\documentclass[review]{elsarticle}

\usepackage{lineno,hyperref,amsmath,graphics,dcolumn}
\usepackage{tabularx}
\usepackage{array}

\usepackage{rotating} %landscape tables
\usepackage{booktabs} %pretty tables
\usepackage{multirow} %multirow and multicolumn environments for tables
\usepackage{newtxmath} %per avere lettere greche, in particolare mu, unslanted
\usepackage{subfig}

\graphicspath{{./images/}}
\journal{Atmospheric Environment}

\biboptions{authoryear}

\begin{document}

\begin{frontmatter}

\title{Spatio-temporal modelling of $\text{PM}_{10}$ daily concentrations in Italy using the SPDE approach}

%\author{Guido Fioravanti\fnref{myfootnote}}

\author[ispra]{Guido Fioravanti\corref{mycorrespondingauthor}}
\ead{guido.fioravanti@isprambiente.it}
\cortext[mycorrespondingauthor]{Corresponding author}
\address[ispra]{a)	Italian National Institute for Environmental Protection and Research, Department for environmental evaluation, control and sustainability. Via Vitaliano Brancati 48, 00144 Rome, Italy }

\author[ntnu]{Sara Martino}
%\ead{}
\address[ntnu]{Norwegian University of Science and Technology, Trondheim, Norway}

\author[bergamo]{Michela Cameletti}
%\ead{}

\address[bergamo]{University of Bergamo,  Bergamo, Italy}

\author[ispra]{Giorgio Cattani}
%\ead{}

\begin{abstract}
This paper illustrates the main results of a spatio-temporal interpolation process of $\text{PM}_{10}$ concentrations at daily resolution using a set of 410 monitoring sites, distributed throughout the Italian territory, for the year 2015. The interpolation process is based on a Bayesian hierarchical model where the spatial-component is represented through the Stochastic Partial Differential Equation (SPDE) approach with a lag-1 temporal autoregressive component (AR1). Inference is performed through the Integrated Nested Laplace Approximation (INLA). Our model includes 11 spatial and spatio-temporal predictors, including meteorological variables and Aerosol Optical Depth. As the predictors' impact varies across months, the regression is based on 12 monthly models with the same set of covariates. The predictive model performance has been analyzed using a cross-validation study.
Our results show that the predicted and the observed values are well in accordance (correlation range: 0.79 – 0.91; bias: 0.22 – 1.07 $\muup \text{g}/\text{m}^3$; RMSE: 4.9 – 13.9 $\muup \text{g}/\text{m}^3$). 
The model final output is a set of 365 gridded (1km $\times$ 1km) daily $\text{PM}_{10}$ maps over Italy equipped with an uncertainty measure. The spatial prediction performance shows that the interpolation procedure is able to reproduce the large scale data features without unrealistic artifacts in the generated $\text{PM}_{10}$ surfaces. The paper presents also two illustrative examples of practical applications of our model, exceedance probability and population exposure maps.
\end{abstract}


\end{frontmatter}

%\linenumbers

\section{Introduction}

 Worldwide, exposure to single pollutants (such as particulate matter, ozone, nitrogen dioxide) accounts for a large portion of overall mortality and cardio-respiratory morbidity \citep{EEA19}. Accordingly, air pollution is recognized as a major public health issue. Among pollutants, particulate matter (PM) is the one associated most consistently with a variety of adverse health outcomes \citep{Martuzzi_et_al_2006, IARC_2015}, even at very low concentrations \citep{Piscitelli_et_al_2019}. \citet{WHO2013} provides a review of the scientific literature concerning the impacts of air pollutants exposure on human health, while \citet{Samoli_et_al_2013}  investigates the adverse health effect of coarse ($\text{PM}_{10}$) and fine ($\text{PM}_{2.5}$) particulate matter in ten Mediterranean metropolitan areas.
 
 In the European context, Italy sadly boasts some of the worst cities and areas for air pollution. The Po Valley in the North of Italy is one of the largest European regions of particular concern in terms of air quality \citep{Raffaelli_et_al_2020}: high and widespread emissions, along with peculiar orographic and meteorological conditions favour both stagnation and formation of secondary particles in winter, and photochemical smog events in summer \citep{EEA19}. Frequent  $\text{PM}_{10}$ daily limits exceedances are also recorded in south central Italy in the Sacco Valley \citep{ISPRA_2020} and the large Naples-Caserta agglomeration during winter months \citep{DeMarco_et_al_2018}.
 
Over the last decades Italy has recorded an important decrease in pollutant emissions thanks to more stringent measures undertaken in order to meet the targets set by the National Emission Ceilings Directive (Directive 2001/81/EC; \citealp{EU2001}). 
Significant $\text{PM}_{10}$, $\text{PM}_{2.5}$ and $\text{NO}_2$ downward trends have been recorded over large portion of the national monitoring network \citep{ISPRA_2019}. Nonetheless, exceedances of the $\text{PM}_{10}$ daily limit value of 50 $\muup \text{g}/\text{m}^3$ (not to be exceeded more than 35 days a year) and ozone long-term target value of 120 $\muup \text{g}/\text{m}^3$ still remain a problem in many cities and rural areas of the country. 

Understanding how $\text{PM}_{10}$ concentrations vary in both space and time  is fundamental for a proper assessment of population-wide exposure and to formulate appropriate pollution mitigation strategies \citep{Chu_et_al_2015}. While daily resolution for $\text{PM}_{10}$ concentrations is often sufficient for exposure assessments, on the spatial scale, there has been an increasing need of high-resolution  maps on large domains, in order to capture concentrations gradients both on the local and the national scale \citep{Cohen_et_al_2017}.
 To this purpose, spatio-temporal statistical models have rapidly gained  attention in the air quality scientific community \citep{Hoek_2017}. The reason is that, compared to regional scale deterministic models, statistical models are generally easier to implement, require medium sized computing resources and provide higher resolution spatial predictions \citep{Shahraiyni_et_Sodoudi_2016}. 

In the statistical literature, the problem of building spatially continuous concentrations maps over large domains has been approached by different angles. A popular approach is that of Linear Mixed Models (LMM) which combine the possibility to  include complex correlation structures, via easy-to-specify random effects at a low computational cost \citep{Galecki_2013}. LMM can in fact be easily implemented in a frequentist framework, using, amon others, the popular {\tt R} package {\tt nlme} \citep{Pinheiro_et_al_2020}. The use of LMM with regional random effects in the air quality community is reported in recent studies such as 
 \citet{Kloog_et_al_2015} and \citet{Stafoggia_et_al_2017}. One drawback of this methodology  is that spatial dependence is expressed through discrete random effects that are related to geographic defined areas,  resulting in prediction maps with spatial artifact (i.e. slabs), e.g. \citet{Sarafian_et_al_2019} or \citet{Zhang_et_al_2018}. 
 In addition, LMM do not incorporate, in the final product (i.e. the $\text{PM}_{10}$ concentration maps)  the whole uncertainty associated with the unknowns (data, parameters, model structure). A practical air quality management strategy must inform decision makers and stakeholders of such uncertainties, in a straightforward and direct manner \citep{Liu_et_al_2008}.

Bayesian hierarchical models \citep{Clark_and_Gelfand_2006}  are another common approach in air quality studies \citep{Blangiardo_et_al_2019, Huang_et_al_2018, Shaddick_et_al_2017, Forlani_et_al_2020}. This approach allows to model complex phenomena as a hierarchy of simpler sub-models, making it possible to deal with the complexity of spatio-temporal processes in a straightforward way. Covariates as orography or temperature  can be used to explain the large scale variability of the phenomenon under study, while residual dependency can be modelled through a space-time process which  is usually assumed to be a Gaussian Random Field (GRF). 
Moreover, the Bayesian approach allows to  easily take into account the variability related to models and parameters, thus giving a more realistic picture of the uncertainty of the final estimates.

The main drawback is that GRF is hard to deal with when there is a lot of data, making its use for environmental applications on large scale challenging \citep{porcu_book}. 
Most of the studies using hierarchical models with spatial GRF concern relatively small areas such as cities \citep{Pollice_taranto, Sahu_2012} or regions (in the Italian context see for example \citealp{Cameletti_et_al_2011, COCCHI2007, Grisotto_et_al_2016}) or consider large domains but without the temporal component \citep{Beloconi_et_al_2018}.

In addition, the main inferential tool for Bayesian hierarchical models, namely the Markov chain Monte Carlo (MCMC) approach \citep{Gilks_et_al_1995}, despite the existence of user friendly programming tools like WinBUGS \citep{BUGS}, JAGS \citep{jags} and  Stan \citep{stan}, can be viewed by the applied community as rather cumbersome, requiring a lot of CPU-time as well as tweaking of simulation and model parameters' specifications. 

Some strategies have been proposed to alleviate the computational burden  of fitting complex spatio-temporal hierarchical models (\citet{Heaton_et_al} for an updated review). One of such strategies, the so-called SPDE (Stochastic Partial Differential Equation) approach, has received a lot of attention in recent years (see \citealp{Bakka_et_al_2018} and reference therein). The SPDE approach provides a way to represent a continuous GRF through a discretely indexed Gaussian Markov Random Field (GMRF; \citealp{lindgren_spde}). Computationally, GMRFs are much more efficient as they are based on sparse matrices \citep{rue2005gaussian}. Moreover, GRF with a SPDE representation can be fitted in a Bayesian hierarchical framework using the Integrated Nested Laplace approximation (INLA) approach \citep{Rue_et_al_2009}. INLA is a deterministic method based on approximating the marginal posterior distributions (by using Laplace and other numerical approximations and numerical integration schemes) and is usually faster and more accurate than MCMC alternatives. Last, but not least, INLA-SPDE comes with a user friendly {\tt R} implementation, the {\tt r-inla} package. Tutorials and examples are available at the dedicated web site \url{r-inla.org} or in book form (e.g. \citealp{Blangiardo_and_Cameletti_2015,Gomez_2020}). This makes the INLA-SPDE methodology a fast, reliable and easy to use tool also to the practitioners.

 In this paper we tested the INLA-SPDE approach to estimate $\text{PM}_{10}$ daily concentrations on a large space-time domain, namely the entire Italian territory (18 conterminous regions plus two major islands), for one year (2015) based on ground daily $\text{PM}_{10}$ records on ca 400 stations. The final result is a collection of high resolution (1 Km $\times$ 1 Km) daily maps  of $\text{PM}_{10}$ concentrations with an associated measure of uncertainty. Such maps can aid responsible authorities and decision-makers for the development of risk assessment and environmental policies.

 The rest of the paper is organized as follows: in Section \ref{Sec:material_and_methods} we present the input dataset and introduce the statistical model we have chosen to analyse the $\text{PM}_{10}$ concentrations. Section \ref{Sec:results} discusses results, model validation and two possible applications of the model estimates for the assessment of air quality in Italy. We end with conclusions in Section \ref{Sec:conclusion}.

\section{Material and Methods}\label{Sec:material_and_methods}

\subsection{Spatial domain}

The Italian peninsula extends into the Mediterranean sea with a narrow and long shape of about 7500 km of coast line. It includes two large mountain systems (the Alps to the north, and the Apennines which extend north-west to south along the country), a large plain (the Po Valley with a surface of 46000 km$^2$) and two major islands (Sicily and Sardinia). This complex orography leads to a variety of climatic conditions which exert a strong influence on the observed spatial and seasonal variability of pollutants concentrations \citep{Perrino_et_al_2020}.
 
Because of its central position in the Mediterranean Basin, Italy is also affected by periodic Saharan dust events which influence air quality. Multiple studies \citep{Matassoni_et_al_2009, Pey_et_al_2013, Barnaba_et_al_2017, Pikridas_et_al_2018}
have estimated the impact of such events on the yearly average $\text{PM}_{10}$ values in the range 1 - 9 $\muup \text{g}/\text{m}^3$, with a north-to-south positive gradient. There is evidence that this increase in $\text{PM}_{10}$ levels has a further negative impact on human health \citep{Tobias_et_al_2011}.

\subsection{Monitoring sites and concentrations data}\label{Par:inputData}

\begin{figure}
\includegraphics[width=\textwidth]{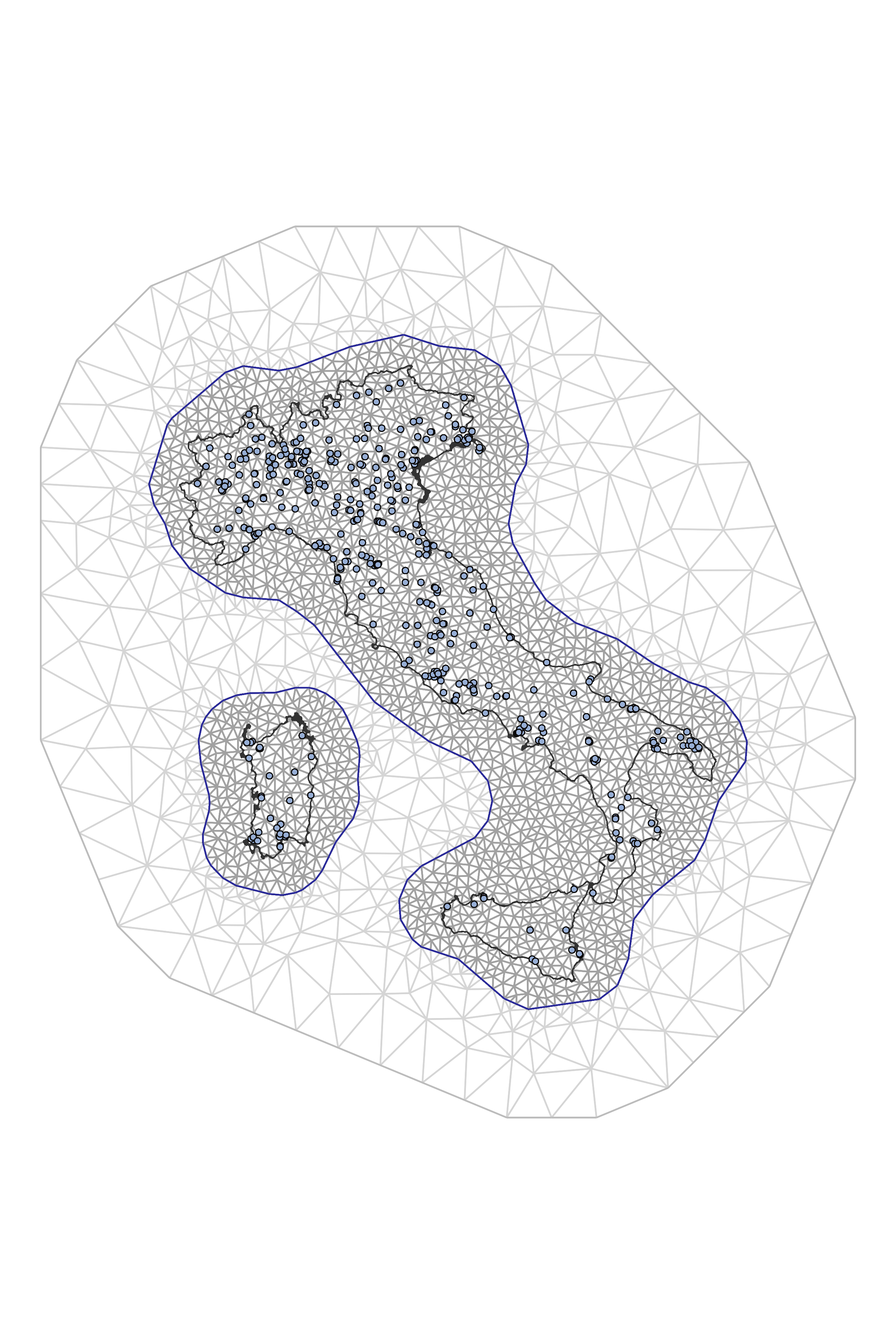}
\caption{Study domain together with the spatial distribution of the 410 monitoring sites. The Figure illustrates also the mesh used to build the SPDE approximation to the continuous Mat\'{e}rn field.}
\label{fig:mesh}
\end{figure}

This study is based on the 2015 $\text{PM}_{10}$ daily average concentrations ($\muup \text{g}/\text{m}^3$) belonging to the Regional Environmental Agencies (ARPA) and collected by the Italian Institute for Environmental Protection and Research (ISPRA). $\text{PM}_{10}$ mass concentrations were determined using the European reference or equivalent methods. The data were fully validated accordingly to standard QA/QC procedures Directive 2008/50/EC \citep{EU2008}.
The data set originally accounted for more than 500 monitoring sites. To work with a more robust dataset, we have removed all stations with less than 10 valid observations per month. The geographical distribution of the final 410 selected stations is shown in Figure \ref{fig:mesh}.  Note that a large portion of the selected time series (83\%) are characterized by low data missingness, having at least 20 observations per month.

The ground $\text{PM}_{10}$ measurements  are mostly located in urban and suburban areas (244 urban stations, 104 suburban and 62 rural). Low elevations are over represented with 75\% of the monitoring sites lying below 250 m. This situation is not unexpected: if on the one hand monitoring networks are designed to ensure observational measures being representative of both high and low polluted areas, on the other hand contaminated areas (located particularly within urban areas) typically require denser networks \citep{EU_2002}.

\begin{figure}
\includegraphics[width=\textwidth]{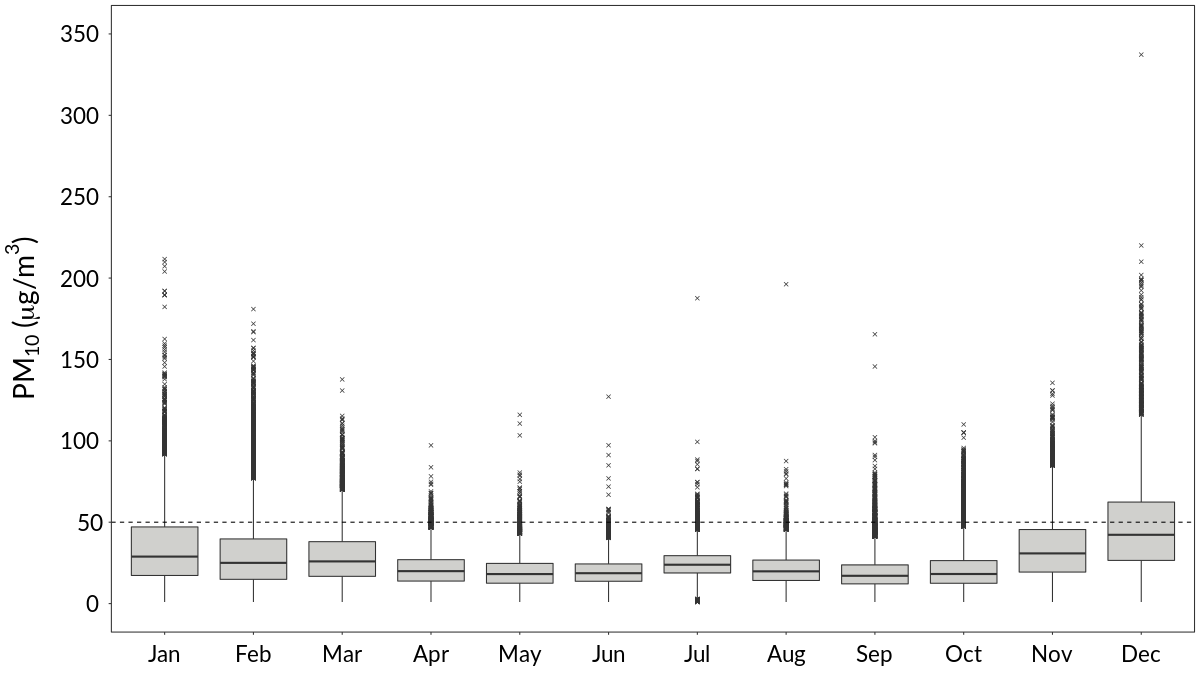}
\caption{Monthly distribution of the daily $\text{PM}_{10}$ concentrations for year 2015. The dashed line indicates the European Community $\text{PM}_{10}$ daily limit value not to be exceeded more than 35 days a year.}\label{fig:tutti}
\end{figure}

The boxplot of  Figure \ref{fig:tutti} shows the $\text{PM}_{10}$ monthly distribution. 
During 2015, $\text{PM}_{10}$ daily concentrations ranged between 0 and 337 $\muup \text{g}/\text{m}^3$, with a median daily  $\text{PM}_{10}$ concentrations of 22.3 $\muup \text{g}/\text{m}^3$ and an inter-quartile range of 15 and 33 $\muup \text{g}/\text{m}^3$. The boxplots suggest a seasonal trend in the observational data: higher $\text{PM}_{10}$ levels, with an average daily median around 30 $\muup \text{g}/\text{m}^3$, characterize the beginning (January-March) and the end (November-December) of 2015. Conversely, lower values were recorded during spring and summer seasons when the average daily median is around 19 $\muup \text{g}/\text{m}^3$. A similar trend characterizes the standard deviation with values around 24 $\muup \text{g}/\text{m}^3$ during the winter months (January-February-December), 14 $\muup \text{g}/\text{m}^3$ during the intermediate seasons (March-April-September-October-November) and 9 $\muup \text{g}/\text{m}^3$ in summer including May.

To conclude this section, we observe that, except for April, all months exhibit occasionally daily values greater than 100 $\muup \text{g}/\text{m}^3$. The three highest values in our input dataset were observed in January (211 $\muup \text{g}/\text{m}^3$), in August (196  $\muup \text{g}/\text{m}^3$) and in December (337 $\muup \text{g}/\text{m}^3$). Despite the outlier nature of these values, the full $\text{PM}_{10}$ distribution was considered and no value was discarded from our analysis. 

\subsection{Predictors}

A large number of potential predictors were available. Based on previous results in the air quality literature, a preliminary analysis of our data, a set of eleven spatial and spatio-temporal predictors was selected to be included in the model by using variable selection methods. The complete list is reported in Table \ref{table:predittori}.

To avoid numerical problems, each predictor (except for the dust indicator) was standardized to have mean 0 and standard deviation 1. All data processing was performed through the combined use of the Climate Data Operator (CDO) software (\url{https://code.mpimet.mpg.de/projects/cdo}), the {\tt R} statistical language \citep{R_CORE_TEAM_2018}  and PostGIS \citep{Strobl_2008}.

In the following, we describe the selected predictors more in details.

\begin{table}\footnotesize
\begin{tabular}{cclp{5.0cm}cc} 
   \toprule
   \textbf{Data Source} & \phantom{abc} & Variable Code &  Description & Unit & Spatial Resolution\\
   \midrule
   \multicolumn{6}{l}{\textbf{ERA5}}\\
    \multicolumn{6}{c}{\phantom{abc}}\\
    \phantom{a} && pbl00 & Planet Boundary Layer at 00:00 & m & \multirow{6}{*}{31 Km}\\ 
    \phantom{a} && pbl12 & Planet Boundary Layer at 12:00 & m & \\ 
    \phantom{a} && ptp & Previous day Total Precipitation & mm & \\        
    \phantom{a} && sp  & Surface Pressure & hPa & \\
    \phantom{a} && t2m & Average temperature at 2 meters & $^{\circ}$C & \\
    \phantom{a} && tp  & Total Precipitation & mm & \\
    \multicolumn{6}{c}{\phantom{abc}}\\
    \multicolumn{6}{l}{\textbf{Copernicus Atmosphere Monitoring Service}}\\
    \multicolumn{6}{c}{\phantom{abc}}\\
    \phantom{a} && aod550   & Aerosol Optical Depth at 550 nm & nm &  $\sim$10 Km\\
    \multicolumn{6}{c}{\phantom{abc}}\\
    \multicolumn{6}{l}{\textbf{Global Multi-resolution Terrain Elevation Data}}\\
    \multicolumn{6}{c}{\phantom{abc}}\\   
    \phantom{a} && q\_dem & Altitude & m &  ~1 Km\\
    \multicolumn{6}{c}{\phantom{abc}}\\
    \multicolumn{6}{l}{\textbf{NMMB-BSC; HYSPLIT-NOAA}}\\
    \multicolumn{6}{c}{\phantom{abc}}\\
    \phantom{a} && dust & Saharan dust & 0/1 &  Macroareas\\
    \multicolumn{6}{c}{\phantom{abc}}\\
    \multicolumn{6}{l}{\textbf{OpenStreetMap}}\\
    \multicolumn{6}{c}{\phantom{abc}}\\
    \phantom{a} && d\_a1 & Linear distance to the nearest highway & m & 1 km\\
    \multicolumn{6}{c}{\phantom{abc}}\\
    \multicolumn{6}{l}{\textbf{Copernicus Land Monitoring Service}}\\
    \multicolumn{6}{c}{\phantom{abc}}\\
    \phantom{a} && i\_surface & Imperviousness & \% & 100 m\\
    \multicolumn{6}{c}{\phantom{abc}}\\
   \bottomrule
\end{tabular}
\caption{List of the predictors included in the spatio-temporal model.}
\label{table:predittori}

\end{table}

\paragraph{Meteorological variables}

Pollutant concentrations are highly dependent on weather conditions \citep{Grange_et_al_2018}, therefore metereological variables are an important part of our model. Hourly surface pressure, total precipitation and temperature at 2 meters height were downloaded as netCDF archives from the ERA5 reanalysis dataset \citep{Hersbach_et_al_2020} of the European Centre for Medium-Range Weather Forecasts (ECMWF). Hourly data were averaged (accumulated, in the case of precipitation) on a daily level. As particulate matter levels depend also on the recent weather history, we have also introduced the variable ``total precipitation of the previous day" \citep{Barmpadimos_et_al_2012}. Planet Boundary Layer height (PBL) is the height up to which the influence of the presence of the lower surface is detectable \citep{Shi_et_al_2020}. PBL at 00:00 and 12:00 was also obtained from the ERA5 dataset and log transformed.

\paragraph{Aerosol Optical Depth}

Aerosol Optical Depth is a key parameter to measure the aerosol “column burden” \citep{Hidy_et_al_2009}. Namely, it represents the extinction of the solar radiation in the atmospheric column attributed to aerosols \citep{Segura_et_al_2017}. $\text{PM}_{10}$ has been shown to correlate with Aerosol Optical Depth \citep{Di_et_al_2016}. In this study, we used numerically simulated estimates of AOD data at the wavelength of 550 nm from the CAMS reanalysis (Copernicus Atmosphere Monitoring Service), whose horizontal spatial resolution is of about 10 kms. The interesting aspect of such data is that it does not suffer from the presence of non-random missing values, which typically affect the well-known satellite product AOD from the Multi-Angle Implementation of Atmospheric Correction (MAIAC) algorithm \citep{Lyapustin_et_al_2018}. 

\paragraph{Elevation}

Elevation data were retrieved from the Global Multi-resolution Terrain Elevation Data of the USGS \citep{Danielson_and_Gesch_2011} at a 30-arc-second (ca 1Km $\times$ 1Km) resolution. 

\paragraph{Dust events}

In our model, the occurance of dust events is described in terms of a dichotomic variable (dust event/no-dust event). The days with dust events have been identified using simulation models (NMMB/BSC-Dust model; \citealp{Perez_et_al_2011}) and Lagrangian models for the simulation of trajectories (HYSPLIT; \citealp{Stein_et_al_2015}). The final information is available for 5 Italian macro-areas: North, Centre, South, Sicily and Sardinia. 

\paragraph{Road traffic emissions}

Different proxy variables were considered to estimate the impact of road traffic emissions, but only the Euclidean distance from the major roads (highways) entered the final model. The road network data come from the OpenStreetMap project \citep{Haklay_et_al_2010} and were downloaded as .pbf (vector) files from the Geofabrik web service (\url{www.geofabrik.de}). 

\paragraph{Impervious surface}

Imperviousness represents the percentage of soil sealing (the covering of land by an impermeable material). Imperviousness is a key indicator of urbanization which provides an estimation of population distribution \citep{Attarchi_2020}. The degree of imperviousness (0-100\%) was downloaded as a GeoTIFF raster file from the Copernicus Land Monitoring Service \citep{Langanke_2016}. 

\subsection{Statistical Modeling}\label{Sec:model}

Let  $y^m(t,s_i)$  denote the realization of the space-time process $Y^m(t,s_i)$ that represents the log $\text{PM}_{10}$ concentrations at day $t = 1,\dots, T^m$ of month $m=1,\dots,12$ at location $s_i$, $i = 1,\dots,410$. The logarithmic transformation is a typical choice for data with highly right skewed distributions \citep{Ott_R_W_1990, Warsono_et_al_2001} like the $\text{PM}_{10}$ data reported in Figure \ref{fig:tutti}.

Our exploratory analysis (results not shown for sake of brevity) highlighted that the impact of each predictor on $\text{PM}_{10}$ concentrations varies across time. Consequently, we developed twelve models, one for each month of the year, all containing the same terms. A similar approach is documented, for example, in \citet{ALHamdan_et_al_2012} for the estimation of $\text{PM}_{2.5}$ concentrations in the Atlanta metropolitan area using AOD data.

We assumed the following model:

\begin{equation}\label{eq:model}
y(t,s_i)  = \mu + \mathbf{x}(t,s_i)\boldsymbol{\beta}^\prime  + u(t,s_i)  +  z(s_i) + \epsilon(t,s_i) 
\end{equation}

Since the models are identical for each month, in the above formula we have omitted the index $m$ to simplify the notation.
In Equation \eqref{eq:model}, $\mu$ is the intercept, $\mathbf{x}(t,s_i) = \left(x_1(t,s_i),\dots,x_p(t,s_i) \right)$ denotes the vector of predictors at site $s_i$ in day $t$ (see Table \ref{table:predittori}) and $\boldsymbol{\beta}=\left(\beta_1,\ldots,\beta_p\right)$ is the corresponding coefficients vector. The term $\epsilon(t,s_i)$ represents measurement error and is defined by a Gaussian white noise process independent over space and time with standard deviation $\sigma_\epsilon$. The process $u(t,s_i)$ represents the residual space-time correlation once the large scale component $\mathbf{x}(t,s_i)\boldsymbol{\beta}^\prime$ is taken into account. As particulate levels are characterized by inter-daily correlation, we assumed $u(t, s_i)$ to change in time according to a first order autoregressive process with spatially colored innovations:
\[ u(t,s_i)=a\; u(t-1,s_i)+\omega(t,s_i) \]
for $t = 2;\dots,T-1$, $|a|<1$. We assumed the innovation $\omega(t,s_i)$ to be a Gaussian process with  mean 0 and covariance  function given by:
\begin{equation}\label{eq:ar1}
\text{Cov}(\omega(t,s_i), \omega(t',s_j)) = 
\left\{\begin{array}{lr}
        0, & \text{for } t\neq t'\\
        \mathcal{C}(h), & \text{for } t = t'\\
        \end{array}\right.
\end{equation}
where $h =||s_i-s_j||$ is the Euclidean distance between sites $i$ and $j$. A common specification for the purely spatial covariance function  $\mathcal{C}(h)$ is the Mat\'{e}rn function:
\[
\mathcal{C}(h) = \sigma^2_{\omega}\frac{1}{\Gamma(\nu)2^{\nu-1}}(k\ h)^\nu K_\nu(k\ h)
\]
where $\sigma^2_{\omega}$ is the marginal variance of the process and $K_\nu(\cdot)$ denotes the Bessel function of second kind and  order $\nu>0$. The parameter $\nu$ measures the degree of spatial smoothness of the process. This parameter is hard to estimate and is usually fixed to a given value rather than estimated, with $\nu = 1$ a common choice \citep{Blangiardo_and_Cameletti_2015}.  The term $k>0$ is a scaling parameter related to the range $\rho$, i.e. the distance at which the spatial correlation becomes small. Following \citet{lindgren_spde},  we used the empirically derived definition $\rho = \frac{\sqrt{8\nu}}{k}$ , with $\rho$ corresponding to the distance where the spatial correlation is close to 0.1, for each $\nu$. To represent the continuous field $u(t, s_i)$  as a GMRF, we used the SPDE approach \citep{lindgren_spde}, which is based on  the finite element method (fem). The triangulation used for fem  in our case is shown in Figure \ref{fig:mesh}. In order to obtain accurate approximations of the underlying  GRF, the triangular mesh must be dense enough to capture the  spatial variability of daily $\text{PM}_{10}$. It is noteworthy to observe that we constructed a mesh which is rather dense over areas with observations and sparser in the outer region, where no data are observed and where we are not interested in prediction. The purpose of the outer mesh is to avoid boundary effects and its sparse triangulation allows to reduce computational costs.

Finally,  the last term in Equation \eqref{eq:model} is defined as $z(s_i)\sim N(0,\sigma^2_z)$  and  is a spatially uncorrelated Gaussian random effect which captures some of the small scale spatial variability. 

\subsection{Priors definition}

In a Bayesian context, in order to finalize the model we need to define prior distributions for the vector $\boldsymbol{\beta}$, the standard deviations $\sigma_\epsilon, \sigma_z$, $\sigma_\omega$, the autocorrelation parameter $a$ in Equation \eqref{eq:ar1}  and the range  $\rho$ of the Mat\'{e}rn function. We used vague Gaussian  priors for the elements of $\boldsymbol{\beta}$ and Penalized Complexity (PC) priors \citep{simpson2017} for the other parameters. The latter are designed to penalize model complexity and avoid overfitting. 
PC priors for the standard deviation parameters can be defined through $\text{Prob}(\sigma>u_\sigma) = \alpha_\sigma$ where $u_\sigma>0$ is a quantile of the prior and $0\leq\alpha_\sigma\leq 1$ is a probability value. In our study we set $u_\sigma = 1$ and $\alpha_\sigma = 0.01$ for both $\sigma_\epsilon, \sigma_z$. The choice was motivated by the fact that the total standard deviation of the observed log $\text{PM}_{10}$ values varies between 0.4 and 0.8 depending on the month, therefore it is very likely for the  variance of each component to be less than 1.  
For $\rho$ and $\sigma_\omega$ we used the joint PC prior suggested in \citet{Fuglstad2019} which can be specified through
\[\text{Prob}(\rho < u_\rho) = \alpha_\rho;\text{  } \text{Prob}(\sigma_\omega > u_{\sigma_\omega}) = \alpha_{\sigma_\omega},\] 
where we set $u_\rho = 150$, $\alpha_\rho = 0.8$, $u_{\sigma_\omega} = 1$, $\alpha_{\sigma_\omega} = 0.01$. Since the large scale spatial dependence is explained by the covariates, it is reasonable to assume the range of the innovation process to be smaller  than 150 Km.
Finally, for the autocorrelation parameter $a$ we used the PC prior proposed in \citet{Sorbye_ar1}. This can be specified through $\text{Prob}(a>u_a)=\alpha_a$, where we set $u_a = 0.8$ and $\alpha_a = 0.4$. The choice was guided by previous findings (e.g. \citealp{Cameletti_et_al_2013}) and restrictions to the possible values of $u_a$ and $\alpha_a$. 

\section{Results and discussion}\label{Sec:results}
Data analysis and modeling have been performed  using the {\tt R} software and in particular the {\tt r-inla} package. Excerpts of the {\tt R} code for the definition of the PC priors and the model fit are available at  \url{https://github.com/guidofioravanti/spde_spatio_temporal_pm10_modelling_italy}. 

 In this section we first discuss parameter estimates and residual analysis for the 12 monthly models. We then show a cross-validation study aimed at assessing the  model performance. Finally,  we present some additional outcomes based on the $\text{PM}_{10}$ spatial predictors available for the 1Km $\times$ 1Km grid covering the whole Italian territory. 
 
\subsection{Parameter estimates}

\begin{figure}
\includegraphics[width=\textwidth]{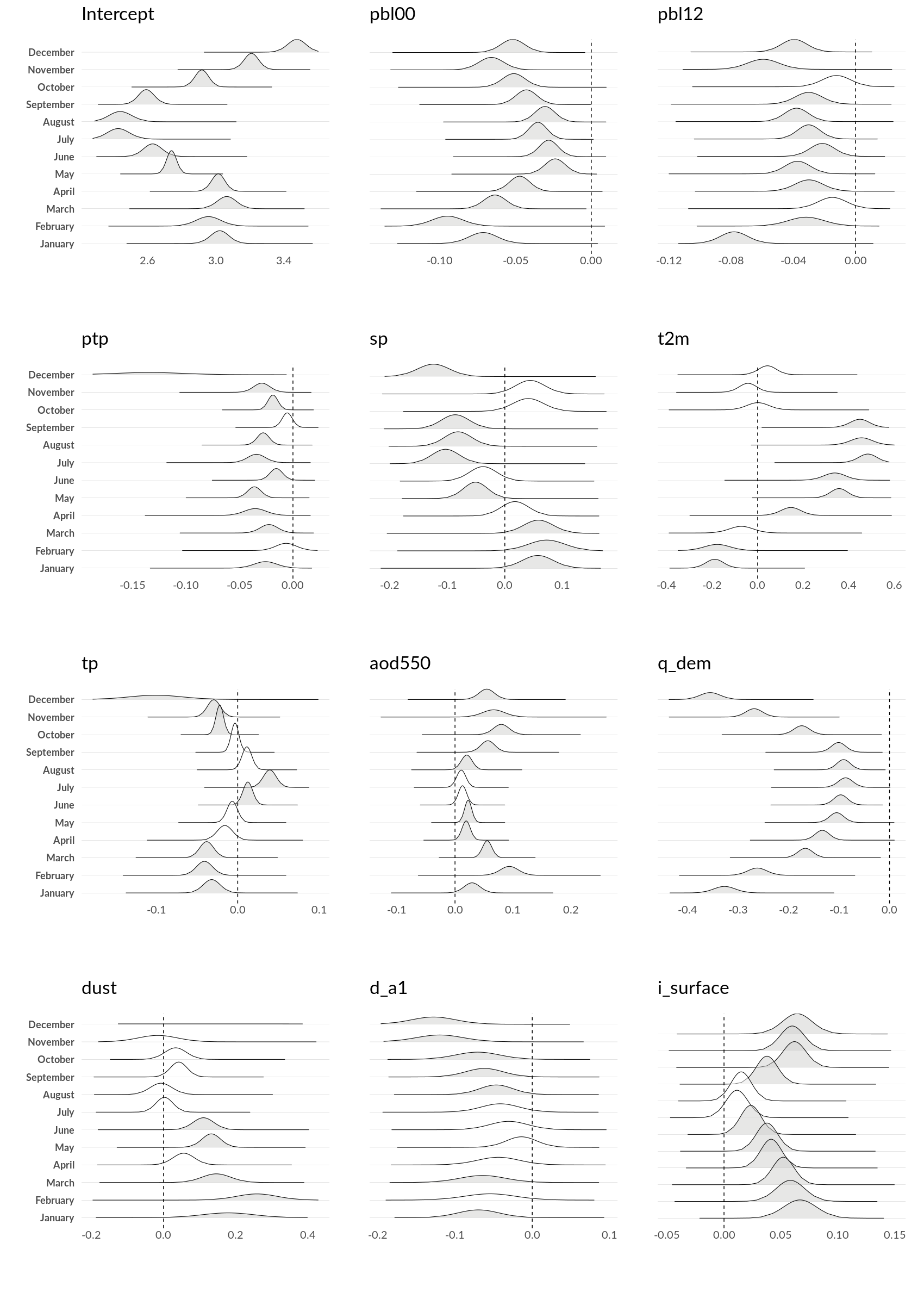}
\caption{Posterior distribution of the intercept $\mu$ and covariate coefficients $\boldsymbol{\beta}$. The shaded color indicates a statistically significant effect.}
\label{fig:postTutte}
\end{figure}

Figure \ref{fig:postTutte} illustrates the posterior distribution for the model intercept $\mu$ and  the 11 covariate coefficients $\boldsymbol{\beta}$ for each of the 12 monthly models.

As expected, many of the  parameters show a clear seasonal behaviour. The posterior mean of  $\mu$ varies from a minimum of 2.42 in July to a maximum of 3.4 in December on the log scale. This corresponds to an average pollution level that varies between 11.2 and 40.0 $\muup \text{g}/\text{m}^3$, after adjustment for covariates. 

The predictors with the most pronounced seasonal effect are: temperature (t2m), Planet Boundary Layer at 00:00 (pbl00), altitude (q\_dem) and impervious surface (i\_surface). Temperature tends to have a positive effect during the summer months and a negative or null effect during the winter months; pbl00 and altitude have negative effects on the log $\text{PM}_{10}$ concentrations, with a stronger magnitude in the winter season. Conversely, the impervious surface has a positive effect, which also tends to be larger in winter. 

%grafico covariate, marginal posterior distribution

 In general, all the covariates, including AOD, have a stronger effect in winter time, when the $\text{PM}_{10}$ levels are higher and more variable both in space and time. Interestingly, we point out that a seasonal effect of the AOD  has  been reported also in \citet{ALHamdan_et_al_2012}, but of opposite sign (weaker during the cool season and relatively strong during the warm season). 
 
The seasonal-varying effects shown in Figure \ref{fig:postTutte} support our initial hypothesis that a monthly regression analysis could improve the accuracy of the final estimates \citep{Weber_et_al_2010}. 

The posterior standard deviation (sd) of the $\boldsymbol{\beta}$ parameters (which can be inferred from the shape of the posterior distributions in Figure \ref{fig:postTutte}) is rather stable from month to month. Exceptions are the sd for the distribution  of the dust indicator and  the total precipitation (same and previous day) in December. These standard deviations are much larger that those in the other months, as a result of no-occurrence of dust events and localized and scarce precipitation events in December 2015.

Estimates of the other  model parameters (posterior means and standard deviations) are reported in Table \ref{table:hyper}. We observe that the spatial component shows higher variability than both the measurement error and the spatial unstructured effect. All the three standard deviations have a seasonal variation, being higher in winter than in summer. The spatial range parameter $\rho$ also presents a  variation across months. The posterior mean goes from a minimum of ca 106 Km in January to a maximum of ca 239 Km in August. There is a clear tendency for the range to be larger in summer corresponding to a spatially smoother particulate matter field; the same behaviour holds for the posterior sd.

Finally, the posterior mean of the AR(1) autocorrelation coefficient $a$ oscillates from 0.62 to 0.82 but there is no clear seasonal pattern. The rather high value of the autocorrelation coefficient confirms the presence of short-term persistence of the $\text{PM}_{10}$. 

% Table created by stargazer v.5.2.2 by Marek Hlavac, Harvard University. E-mail: hlavac at fas.harvard.edu
% Date and time: lun, lug 27, 2020 - 21:48:14
\begin{table}[!htbp] \centering 
\footnotesize 
\begin{tabular}{@{\extracolsep{0.5pt}} lccccc} 
\toprule
\phantom{abc} & $a$ & $\rho$ & $\sigma_z$ & $\sigma_{\epsilon}$ & $\sigma_{\omega}$ \\ 
\midrule 
January & 0.629 (0.018) & 106.23 (4.186) & 0.247 (0.012) & 0.197 (0.002) & 0.434 (0.011) \\ 
February & 0.656 (0.017) & 135.213 (5.341) & 0.207 (0.01) & 0.201 (0.002) & 0.513 (0.014) \\ 
March & 0.656 (0.018) & 192.579 (8.498) & 0.162 (0.008) & 0.18 (0.002) & 0.432 (0.014) \\ 
April & 0.742 (0.019) & 153.914 (8.29) & 0.151 (0.007) & 0.178 (0.002) & 0.361 (0.014) \\ 
May & 0.624 (0.02) & 167.292 (9.017) & 0.163 (0.007) & 0.177 (0.002) & 0.289 (0.008) \\ 
June & 0.743 (0.023) & 237.997 (15.118) & 0.157 (0.006) & 0.167 (0.001) & 0.282 (0.013) \\ 
July & 0.823 (0.018) & 177.433 (10.544) & 0.163 (0.007) & 0.155 (0.001) & 0.263 (0.014) \\ 
August & 0.704 (0.022) & 238.714 (15.483) & 0.159 (0.006) & 0.177 (0.001) & 0.256 (0.011) \\ 
September & 0.697 (0.019) & 181.108 (9.8) & 0.161 (0.007) & 0.177 (0.002) & 0.319 (0.011) \\ 
October & 0.727 (0.017) & 164.188 (7.097) & 0.171 (0.007) & 0.176 (0.002) & 0.382 (0.013) \\ 
November & 0.78 (0.014) & 105.514 (3.908) & 0.167 (0.009) & 0.153 (0.002) & 0.443 (0.014) \\ 
December & 0.825 (0.014) & 83.96 (2.968) & 0.209 (0.012) & 0.148 (0.002) & 0.41 (0.015) \\ 
\bottomrule
\end{tabular} 
  \caption{Posterior means (standard deviations) of the  parameters in all 12 models.} 
  \label{table:hyper} 
\end{table} 

\begin{figure}
\includegraphics[width=\textwidth]{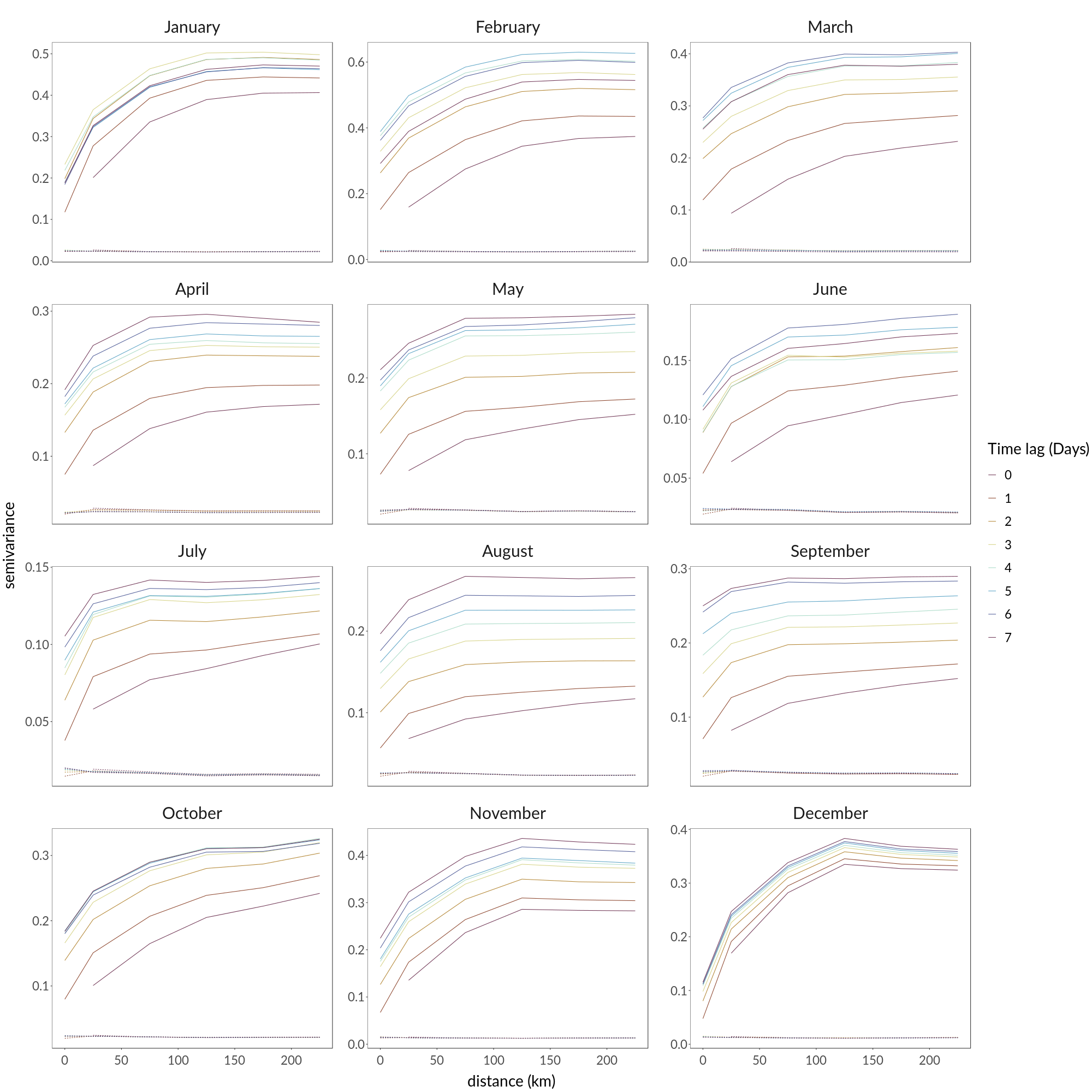}
\caption{Monthly spatio-temporal variograms for the observed (log) $\text{PM}_{10}$ concentrations (solid lines) and the corresponding model residuals (dashed lines).}
\label{fig:variograms}
\end{figure}

In order to assess whether the model manages to capture the spatio-temporal variability of the $\text{PM}_{10}$ observations, we show in Figure \ref{fig:variograms} the spatio-temporal variograms \citep{Cressie_book} 
for the log $\text{PM}_{10}$ concentrations (solid lines) and for the model residuals (dotted lines). 

%grafico variogrammi spazio-temporali

For the log $\text{PM}_{10}$ concentrations the semi-variance increases with distance (x-axis), suggesting spatial dependence among observations. A similar behaviour is apparent when we look at the semi-variance along the y-axis (time), having fixed a distance on the x-axis: in this case, the semi-variance increases with the time-lag, reflecting temporal dependence in the data. None of these patterns can be seen in the corresponding residuals variograms, indicating that the models capture the spatio-temporal signal and return uncorrelated residuals. 

\subsection{Validation}

To evaluate the predictive performance of the model we did a cross-validation study similar to the one presented in \citet{Pirani_et_al_2014}. Specifically, we stratified the 410 input monitoring sites into three groups according to their area type category (urban, suburban and rural). A validation dataset was identified by sampling 10\% of the monitoring sites in each group (24 urban sites, 11 suburban and 6 rural), with the rest of the stations labelled as training dataset. We used the training dataset to fit the model and predict $\text{PM}_{10}$ concentrations on the validation dataset. Finally, we compared the predicted values to the observed ones and summarised the results using a series of performance measures. The sampling process was repeated three times (trials), resulting in three validation and training datasets.

%As performance measures we chose the following indices: the empirical coverage of 95\% credible intervals (95\% CI), the  correlation coefficient, the root mean square error (RMSE) and  the bias (computed as the difference between the observed  concentrations and the posterior predicted mean).

%Table \ref{table:crossval} reports the mean of the performance measures computed over the three different trials. All indices are on the original scale for ease of communication to the practitioners and the end users.

As performance measures we chose the following indices: 1) the empirical coverage of 95\% credible intervals (95\% CI); 2) the correlation coefficient; 3) the root mean square error (RMSE); 4) the bias. The last three indexes are computed by comparing the observed concentrations and the posterior predicted means of each monitoring site. For each training/validation dataset, the average of each performance score over all stations was computed. Table \ref{table:crossval} reports the global model performance in terms of average scores over the three different trials. All indices are on the original scale for ease of communication to the practitioners and the end users.

%tabella cross validation in formato landscape 
\begin{sidewaystable}\footnotesize
\begin{center}
\begin{tabular}{l c c c c c c c c c c c} 
\toprule
\multirow{4}{*}{\phantom{abc}} & \multicolumn{2}{c}{\textbf{RMSE}} & \phantom{a} & \multicolumn{2}{c}{\textbf{Correlation}} & \phantom{a} & \multicolumn{2}{c}{\textbf{Bias}} & \phantom{a} & \multicolumn{2}{c}{\textbf{Coverage}}\\
\phantom{a} & \multicolumn{2}{c}{$\mu g/m^3$} &&  \multicolumn{2}{c}{\phantom{a}} && \multicolumn{2}{c}{$\mu g/m^3$} && \multicolumn{2}{c}{$\%$}\\
\cmidrule{2-3} \cmidrule{5-6} \cmidrule{8-9} \cmidrule{11-12}\\
\phantom{a} & Training & \textit{Validation} && Training & \textit{Validation} && Training & \textit{Validation} && Training & \textit{Validation}\\
\midrule 
\textnormal{January} & 5.33 & \textit{11.14} && 0.98 & \textit{0.87} && -0.04 & \textit{1.07} && 98.23 & \textit{94.93}\\ 
\textnormal{February} & 4.87 & \textit{9.59} && 0.98 & \textit{0.91} && -0.1 & \textit{0.62} && 98.10 & \textit{94.41}\\ 
\textnormal{March} & 4.24 & \textit{7.43} && 0.97 & \textit{0.89} && -0.07 & \textit{0.54} && 97.66 & \textit{94.64}\\ 
\textnormal{April} & 3.08 & \textit{5.60} && 0.95 & \textit{0.83} && -0.03 & \textit{0.49} && 97.61 & \textit{93.74}\\ 
\textnormal{May} & 3.06 & \textit{5.31} && 0.95 & \textit{0.83} && -0.03 & \textit{0.28} && 97.86 & \textit{95.16}\\ 
\textnormal{June} & 3.02 & \textit{4.84} && 0.93 & \textit{0.79} && -0.02 & \textit{0.25} && 97.31 & \textit{94.43}\\ 
\textnormal{July} & 3.47 & \textit{6.38} && 0.92 & \textit{0.71} && -0.01 & \textit{0.29} && 97.41 & \textit{94.56}\\ 
\textnormal{August} & 3.68 & \textit{5.41} && 0.92 & \textit{0.82} && -0.03 & \textit{0.22} && 97.05 & \textit{95.62}\\ 
\textnormal{September} & 3.68 & \textit{5.63} && 0.94 & \textit{0.85} && 0.01 & \textit{0.55} && 97.51 & \textit{95.24}\\ 
\textnormal{October} & 3.21 & \textit{6.05} && 0.97 & \textit{0.89} && -0.01 & \textit{0.61} && 97.80 & \textit{94.61}\\ 
\textnormal{November} & 3.77 & \textit{8.92} && 0.98 & \textit{0.88} && -0.01 & \textit{0.73} && 98.44 & \textit{93.57}\\ 
\textnormal{December} & 5.79 & \textit{13.90} && 0.98 & \textit{0.83} && 0.04 & \textit{0.79} && 98.53 & \textit{95.48}\\ 
\bottomrule 
\end{tabular}
\caption{Statistics of the cross-validation study (on original scale).} 
\label{table:crossval} 
\end{center}
\end{sidewaystable}

Generally speaking, it appears that the models perform well both in the training and in the validation phase. RMSE values are higher in the winter months for both phases. This is not surprising since in winter we observe higher particulate concentrations. 

\begin{figure}%
    \centering
    \subfloat[Training Stage]{\includegraphics[width=1\textwidth]{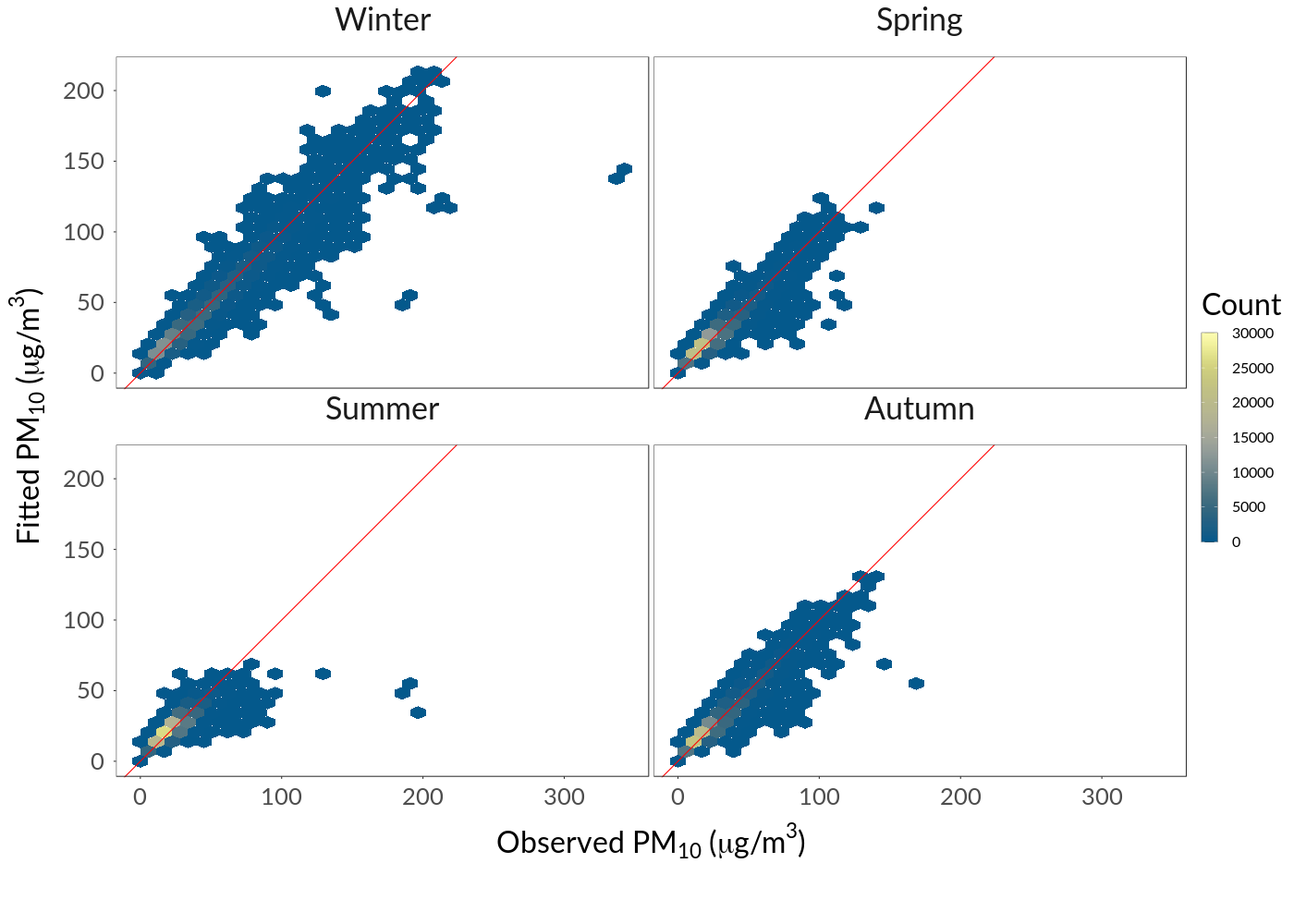}}%

    \subfloat[Validation Stage]{\includegraphics[width=1\textwidth]{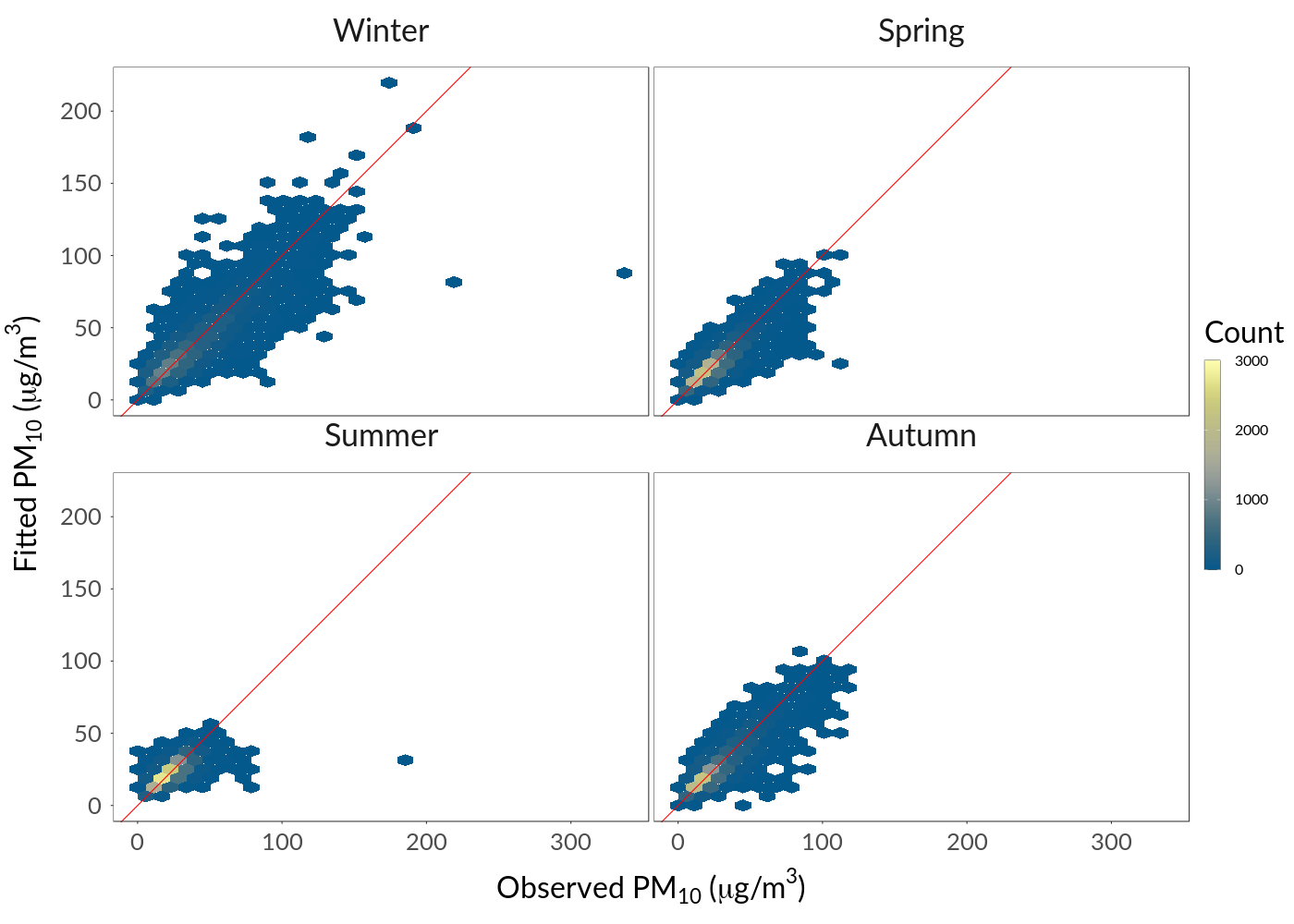}}%
    \caption{Agreement between modelled and measured $\text{PM}_{10}$ concentrations. Lighter colors indicate areas with higher points concentrations. The solid line is the 1:1 line as a reference.}%
    \label{fig:scatterplot}%
\end{figure}

The high values of the correlation coefficients (above 0.9 for all months in the training phase and and above 0.7 in the validation phase) show that the predicted and the observed values are well in accordance. This can be also seen from Figure \ref{fig:scatterplot} where we have plotted the predicted versus the observed values. To avoid having too many scatterplots, in Figure \ref{fig:scatterplot} we adopted a seasonal representation.

The plots highlight that the points are distributed uniformly along the diagonal line. However, a general underestimation of high concentrations values is apparent in all seasons both in the training and validation stage. In particular, we see that the model fails to reproduce very high concentrations above 150 $\muup \text{g}/\text{m}^3$.

Back to Table \ref{table:crossval}, a negligible bias can be observed, with absolute values less than 1.1  $\muup \text{g}/\text{m}^3$ in all months.
Finally, the empirical coverage is very close to its nominal value of 95\%. 

\begin{figure}%
    \centering
    \subfloat[Urban station - January ]{\includegraphics[width=0.45\textwidth]{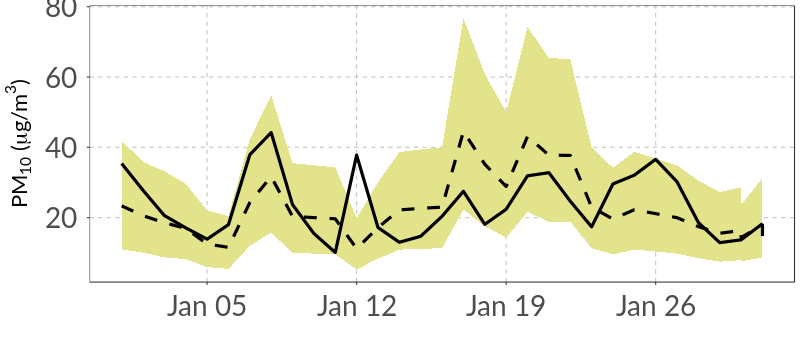}}%
    \qquad
    \subfloat[July]{\includegraphics[width=0.45\textwidth]{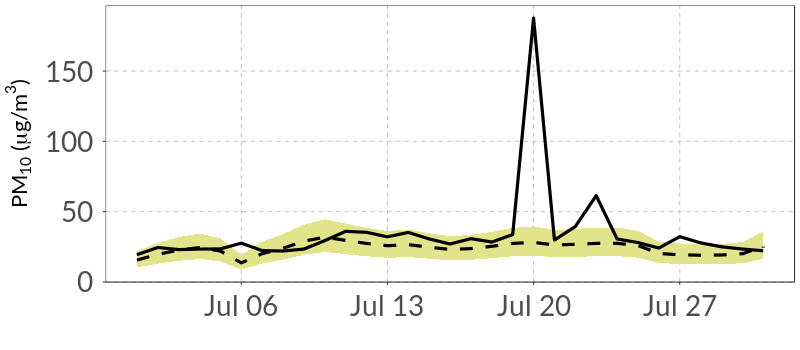}}%
    \qquad
    \subfloat[Suburban station - January ]{\includegraphics[width=0.45\textwidth]{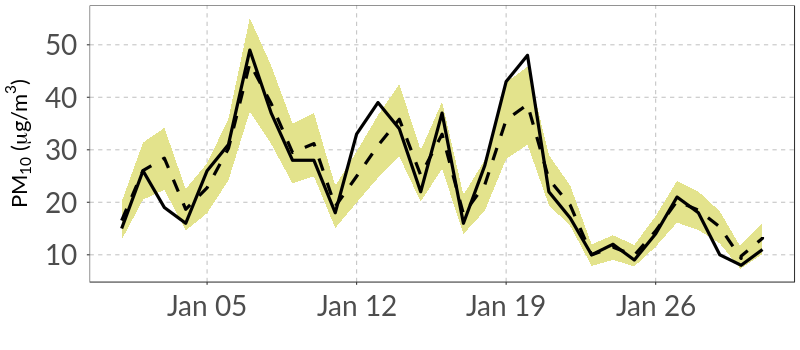}}%
    \qquad
    \subfloat[July]{\includegraphics[width=0.45\textwidth]{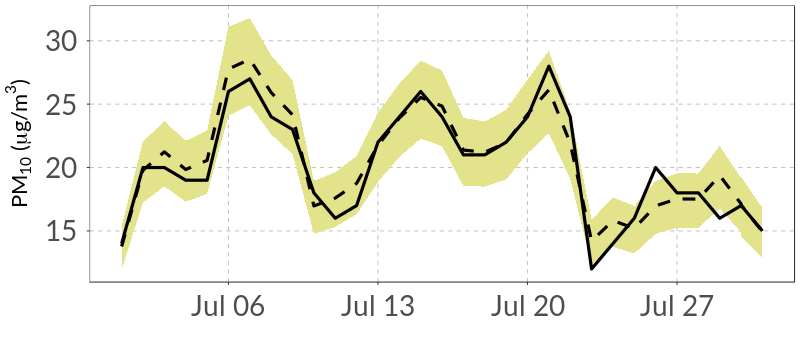}}%    
    \qquad
    \subfloat[Rural station - January ]{\includegraphics[width=0.45\textwidth]{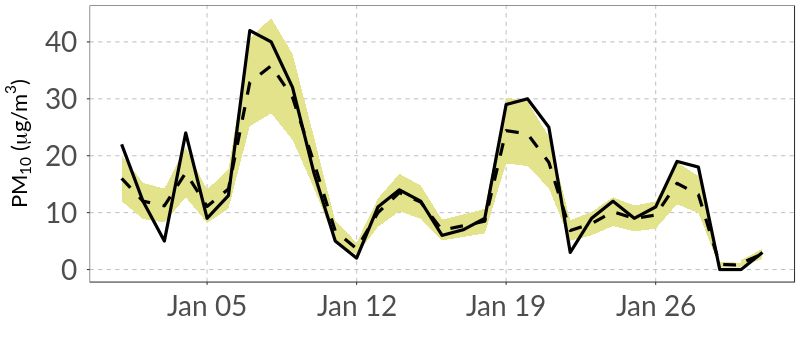}}%
    \qquad
    \subfloat[July]{\includegraphics[width=0.45\textwidth]{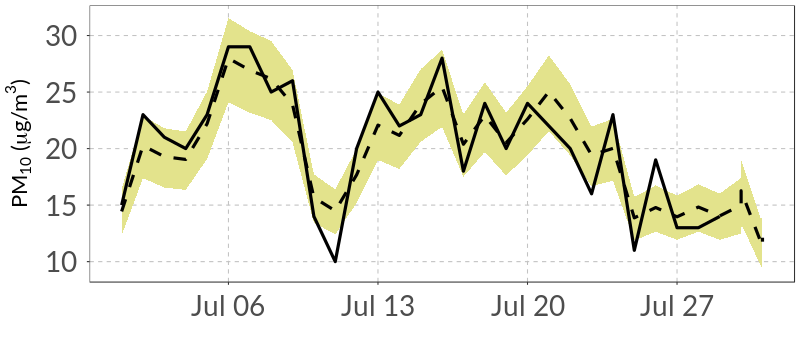}}    
    \caption{$\text{PM}_{10}$ daily concentrations for three illustrative monitoring sites, one for each type area category (urban, suburban and rural). Observed (solid lines) versus fitted values (dashed lines).}%
    \label{fig:timeseries}%
\end{figure}

Figure \ref{fig:timeseries} shows a comparison between observed and predicted time series for 3 illustrative stations chosen from the validation set. For sake of brevity, we present the results for two months alone: January and July. The time series plots suggest that the model is able to reproduce the temporal variability of the monitoring sites in the validation dataset, although some very high values (for example in the upper right panel of Figure \ref{fig:timeseries}) are not properly captured. 

%serie temporali, trial 3
%\begin{figure}
%\includegraphics[width=\textwidth]{images/graficiSerieValidazione14_trail3.png}
%\caption{$\text{PM}_{10}$ daily concentrations: observed (in black) and fitted values (in red). The left and right panels display the series for the months of January and July, respectively.}
%\label{fig:timeseries}
%\end{figure}

\subsection{Spatial Prediction}\label{Sec:SpatialPrediction}

In this section, we focus on spatial predictions. In particular, we provide examples of daily and monthly maps, using a 1km $\times$ 1km grid over the whole Italian territory. This results in a spatial grid of 310622 cells which, across the entire year 2015, corresponds to a spatio-temporal grid of over 11 millions cells.

We have simulated 1000 samples from the posterior distribution of all model components for two months. We chose January and July 2015 in  order to show some of the seasonal characteristics of the fitted model. Having a sample distribution of 1000 gridded maps for each day of January and July 2015, we were able to calculate summary statistics of central tendency (mean) and variability (sd). 

%grafico mappe giornaliere media
%\begin{figure}
%\includegraphics[width=\textwidth]{images/grafico_medie_rocv_giornaliere.png}
%\caption{Posterior daily mean $\text{PM}_{10}$ concentrations maps (upper plots) and relative width of the posterior interquartile range (lower plots).}
%\label{fig:mappeGiornaliere_media}
%\end{figure}

\begin{figure}%
    \centering
    \subfloat[January 26th]{\includegraphics[width=0.45\textwidth]{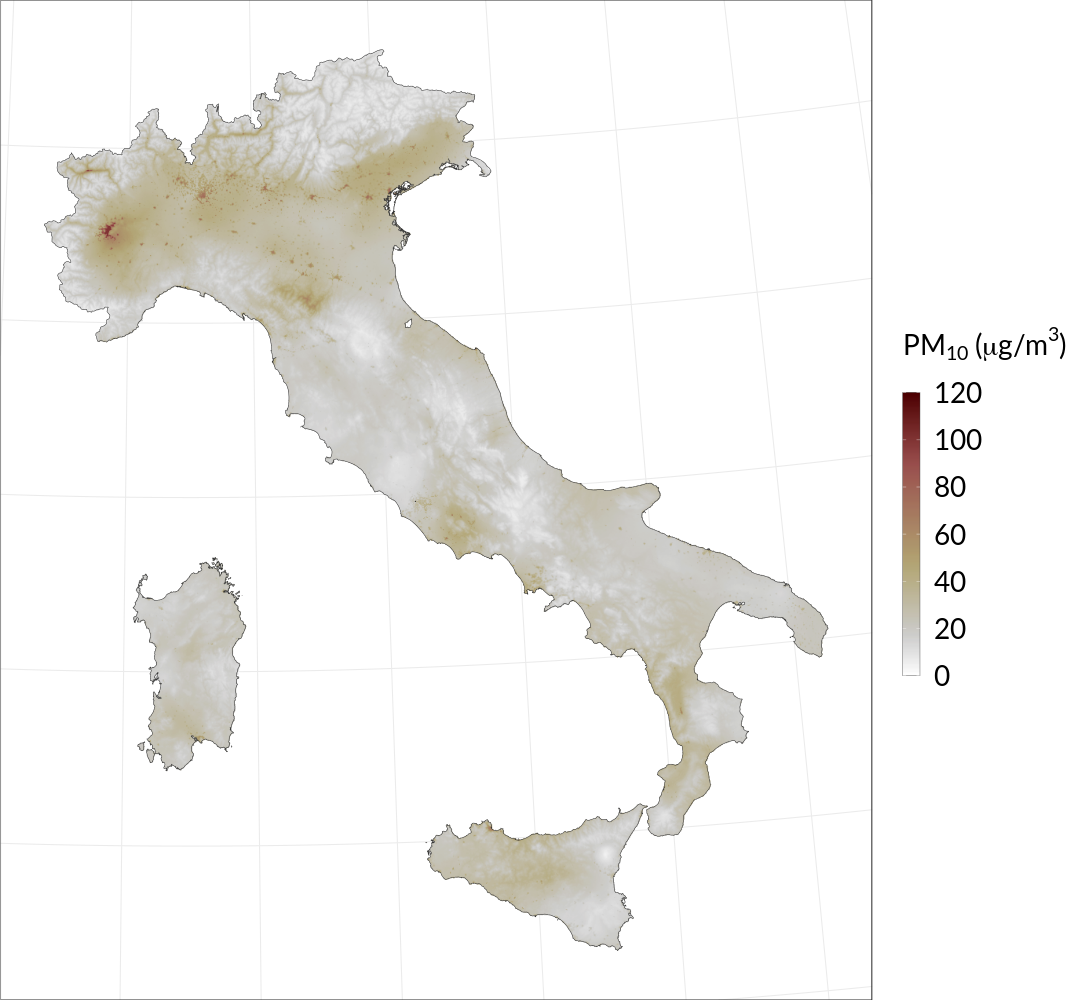}}%
    \qquad
    \subfloat[July 21st]{\includegraphics[width=0.45\textwidth]{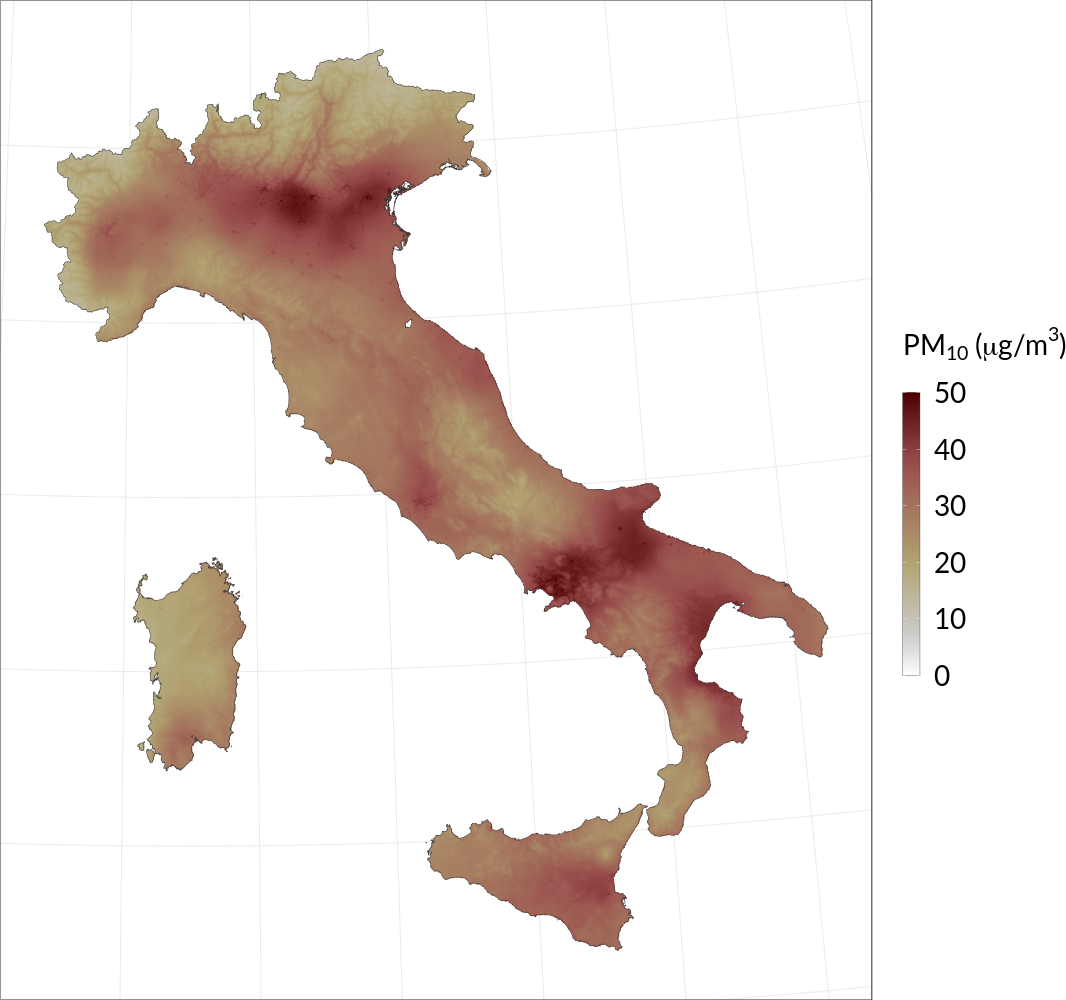}}%
    \qquad
    \subfloat[January 26th]{\includegraphics[width=0.45\textwidth]{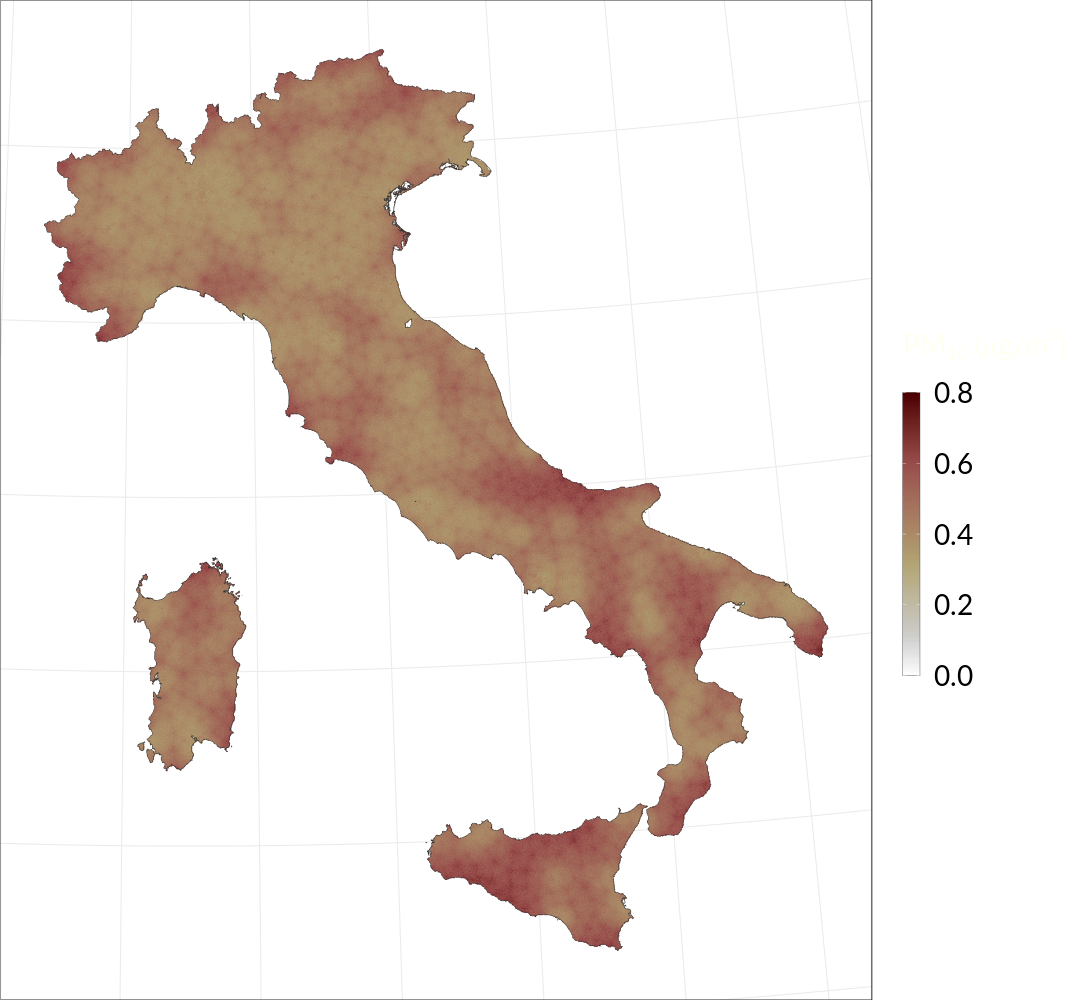}}%
    \qquad
    \subfloat[July 21st]{\includegraphics[width=0.45\textwidth]{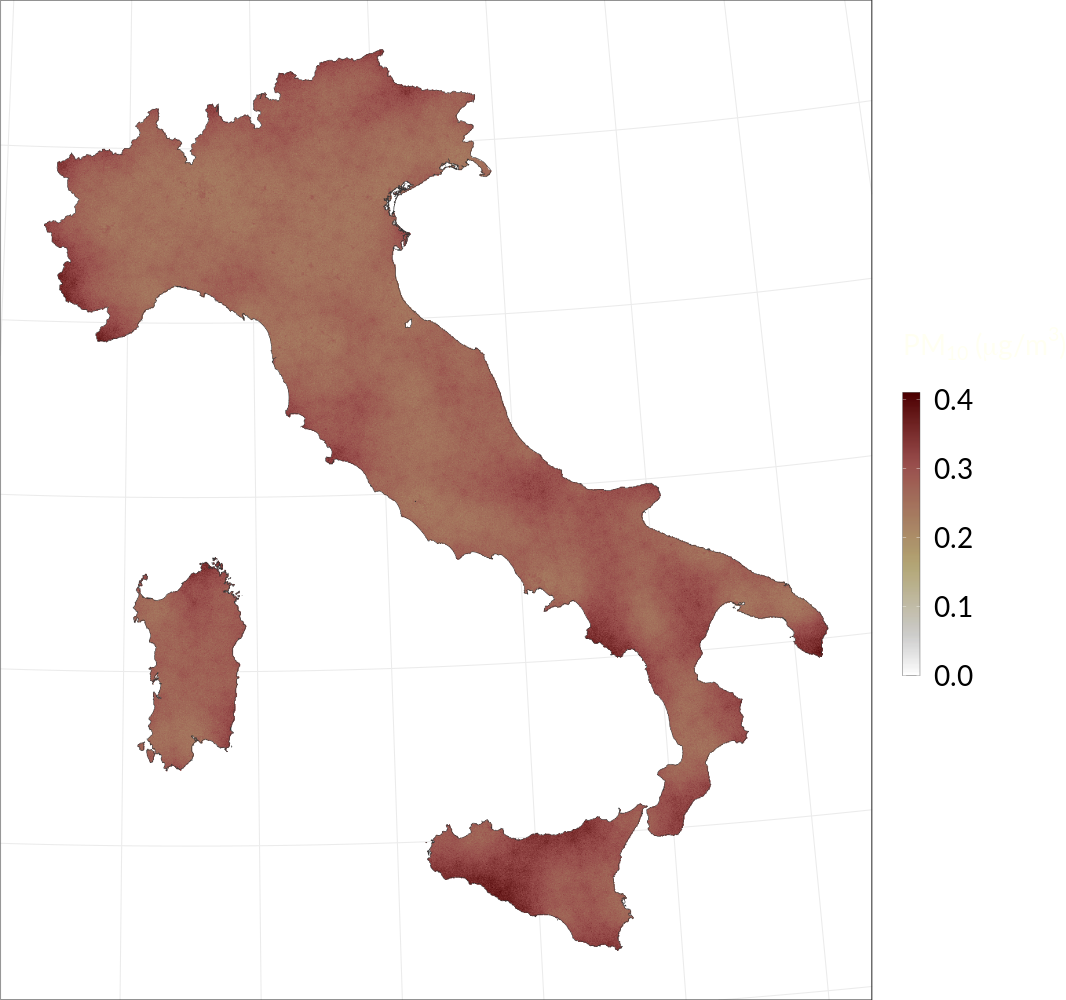}}%    
    \caption{Posterior daily mean $\text{PM}_{10}$ concentrations maps (a-b) and relative width of the posterior interquartile range (c-d) for January 26th and July 21st, 2015.}%
    \label{fig:mappeGiornaliere_media}%
\end{figure}

As an example, Figure \ref{fig:mappeGiornaliere_media} a) and b) show the posterior mean of the daily $\text{PM}_{10}$ concentrations on January 26th and July 21st 2015. These two dates were chosen randomly and have no special meaning. Note that the two figures have different color scales. 
A visual inspection of Figure \ref{fig:mappeGiornaliere_media} a) and b) highlights that the interpolation procedure is able to reproduce the large-scale data features without unrealistic artifacts in the generated surfaces. Specifically, both daily maps exhibit a reasonable spatial pattern of high $\text{PM}_{10}$ mean concentrations in urbanized environments, which decrease in rural areas and with altitude. This is especially apparent in the January map, when the model estimates high $\text{PM}_{10}$ levels in the Po Valley with a peak above 50 $\muup \text{g}/\text{m}^3$ in the Turin city area (North-Western Italy). In July, the model generates a smoother surface with less spatial variability. The orography here, for example, is visible but less pronounced than in January. This result is not unanticipated: it reflects the results seen in Table \ref{table:hyper}, the greater range and lower variability of the latent spatial field in summer with respect to the winter time. These results, in turn, depend on the seasonality of the $\text{PM}_{10}$ concentrations illustrated through the boxplots of Figure \ref{fig:tutti}. 

A video, describing the entire temporal evolution of the daily $\text{PM}_{10}$ concentrations for both months of January and July 2015 is available at \url{https://github.com/guidofioravanti/spde_spatio_temporal_pm10_modelling_italy}.

We use the relative width of the posterior interquartile range (RWPIR) as a measure for the relative uncertainty of the predicted concentrations surface \citep{yuan2017}:
\[
RWPIR = (Q_3-Q_1)/Q_2,
\]
where $Q_1, Q_2$ and $Q_3$ are the first quartile, the median and the third quartile.

The RWPIR for the two selected days is shown in Figure \ref{fig:mappeGiornaliere_media} c) and d) for January 26th and June 21st, respectively. As expected, the relative uncertainty is higher in January than in July but the spatial pattern in Figure \ref{fig:mappeGiornaliere_media} c) and d) is quite similar: uncertainty is lower where there are more monitoring sites and higher otherwise. 

\begin{figure}%
    \centering
    \subfloat[January]{\includegraphics[width=0.45\textwidth]{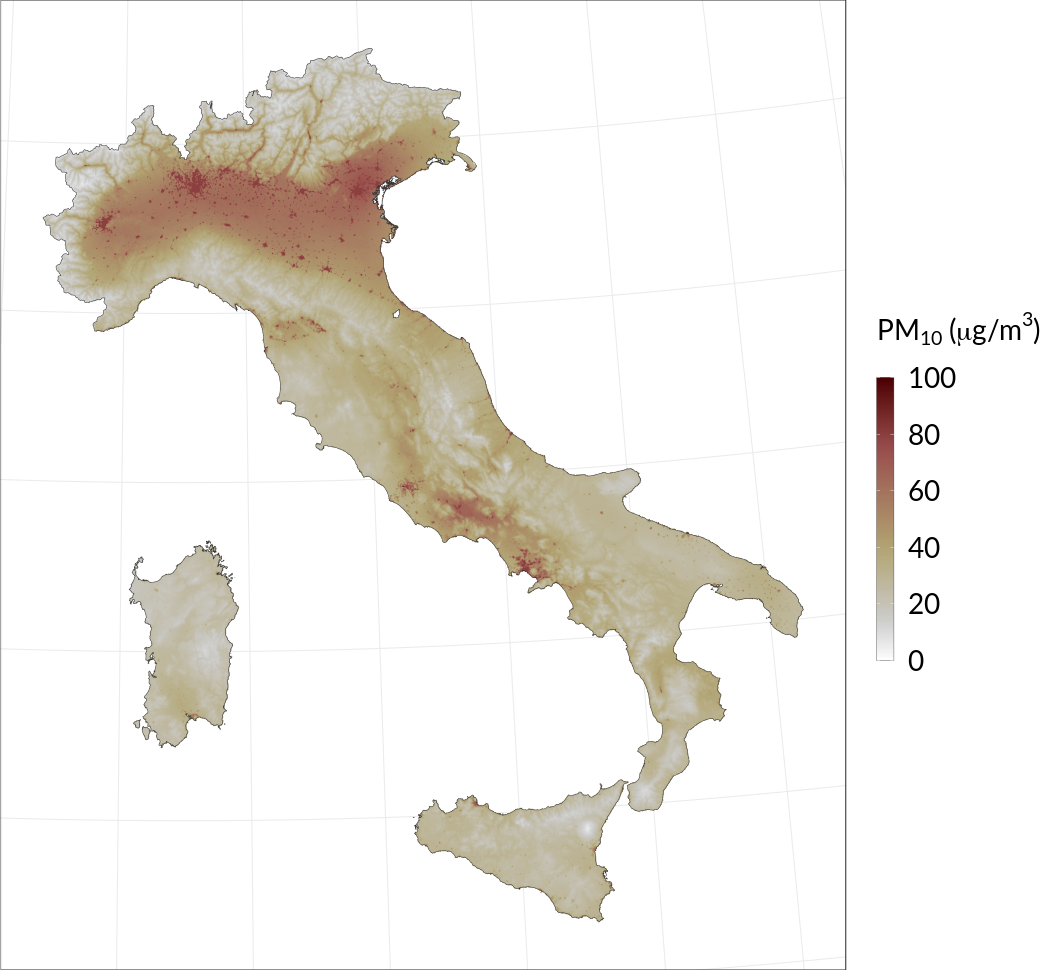}}%
    \qquad
    \subfloat[July]{\includegraphics[width=0.45\textwidth]{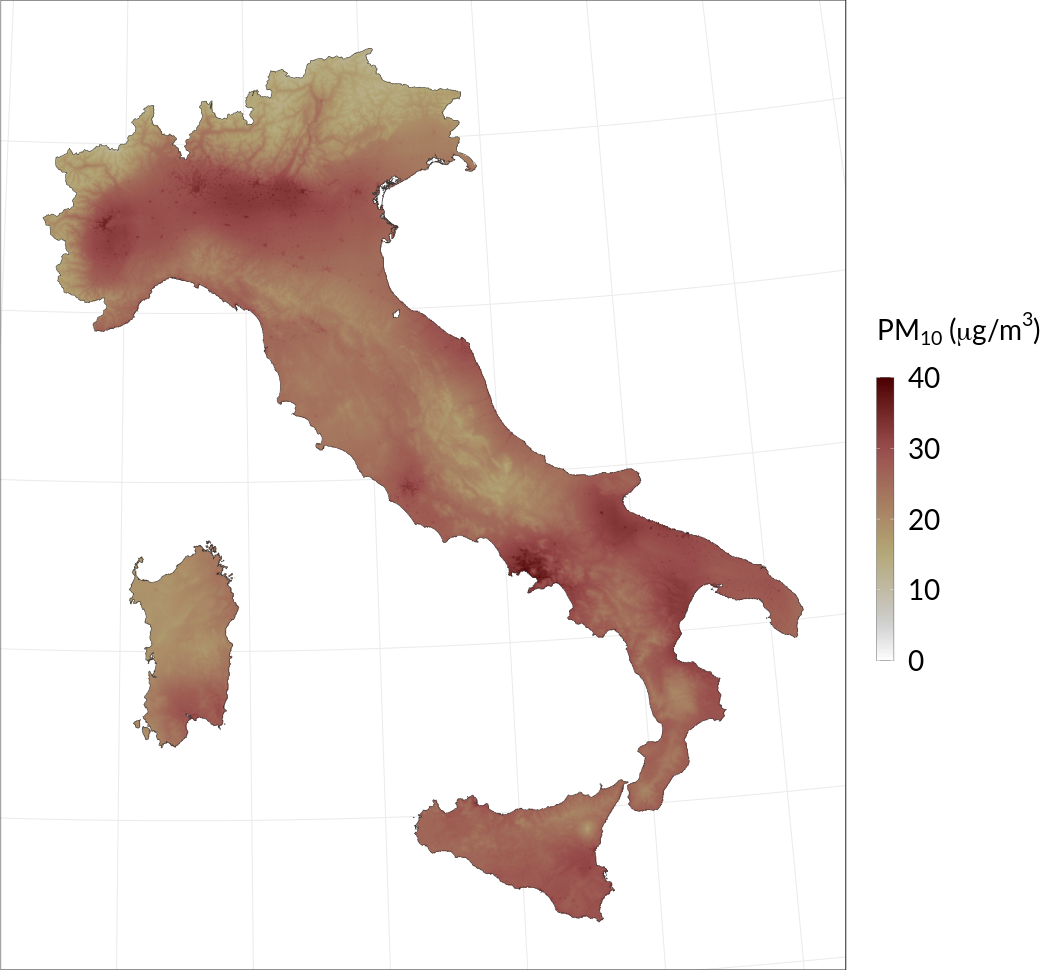}}%
    \qquad
    \subfloat[January]{\includegraphics[width=0.45\textwidth]{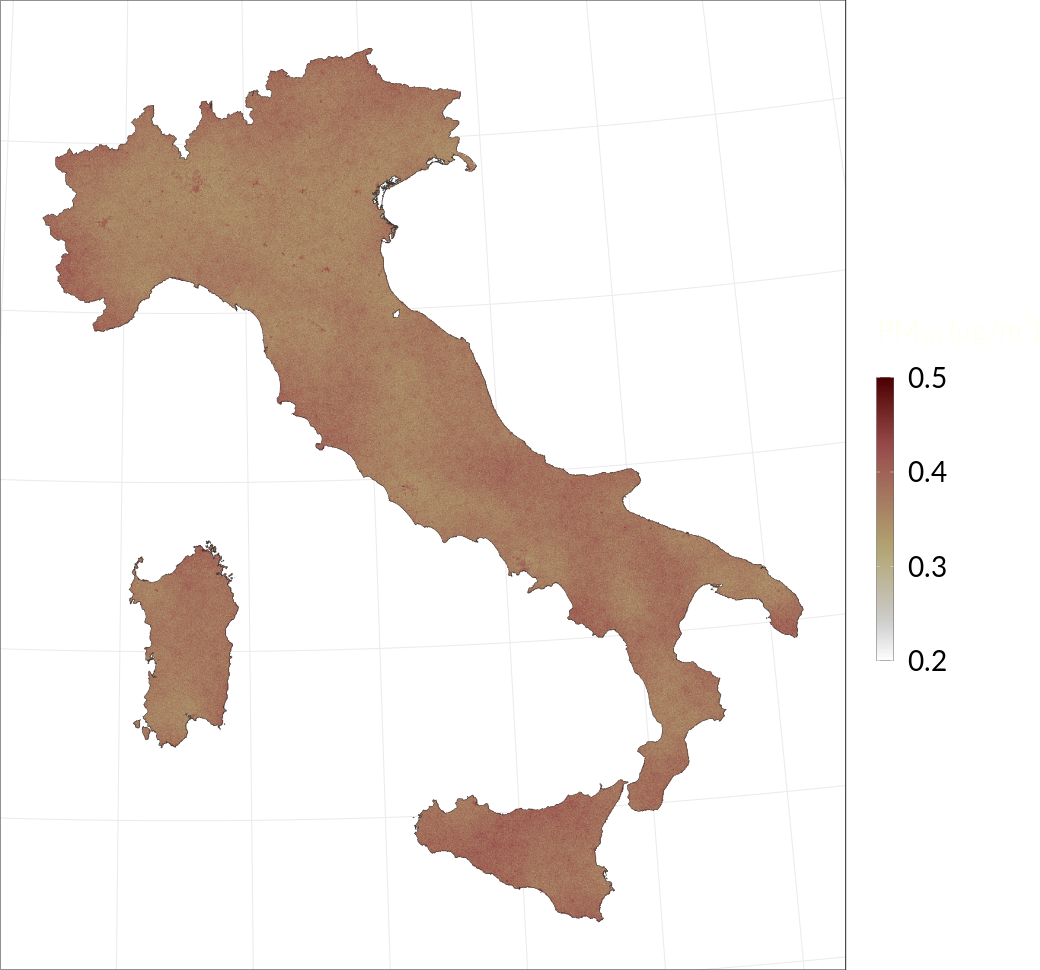}}%
    \qquad
    \subfloat[July]{\includegraphics[width=0.45\textwidth]{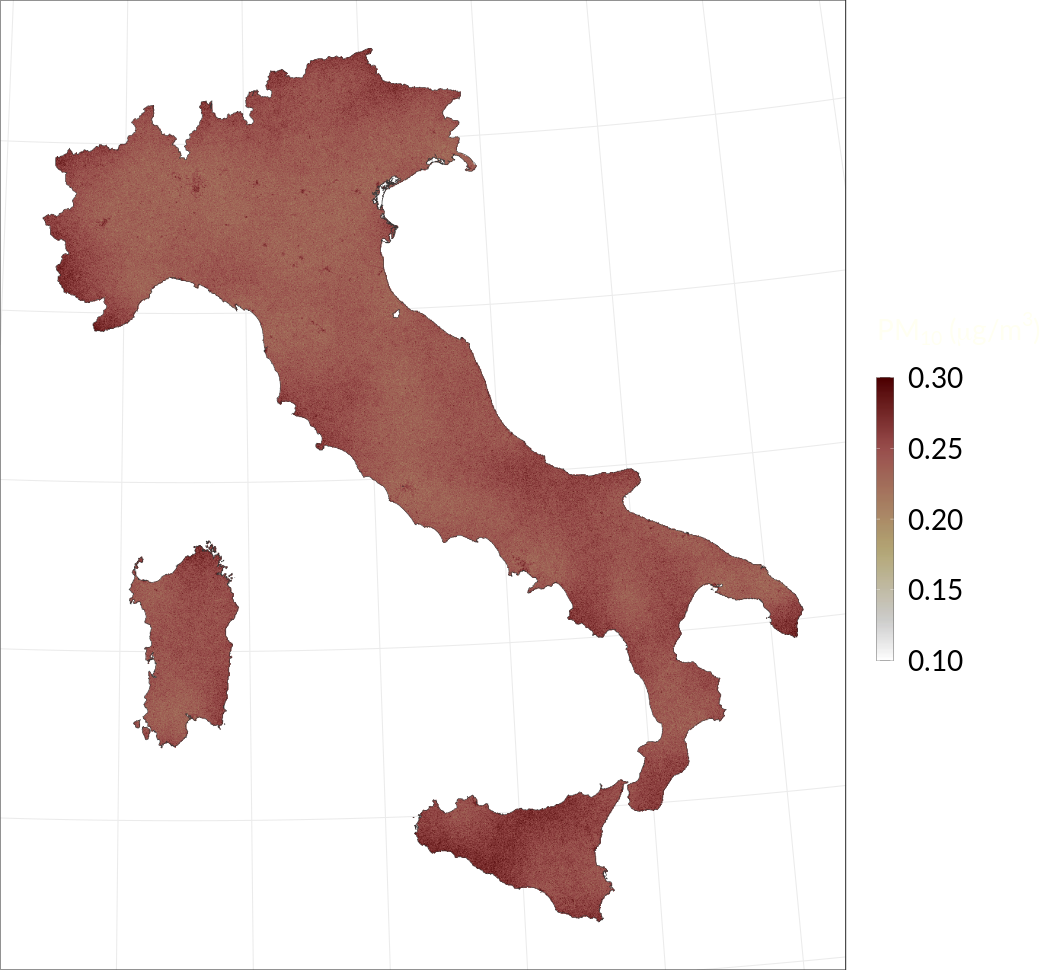}}%    
    \caption{Monthly average $\text{PM}_{10}$ concentrations maps (a-b) and relative width of the posterior interquartile range (c-d) for January and July 2015.}%
    \label{fig:mappeMensili_media}%
\end{figure}

Analogous considerations apply when we examine the monthly average concentrations maps.  Figure \ref{fig:mappeMensili_media} a) and b) show the posterior monthly PM$_{10}$ average concentrations while Figure  \ref{fig:mappeMensili_media} c) and d) shows the RWPIR. In this case, the simulated daily prediction surfaces were aggregated in order to create a corresponding sample of 1000 average monthly concentrations maps.

%grafico mappe mensili media
%\begin{figure}
%\includegraphics[width=\textwidth]{images/grafico_medie_rocv_mensili.png}
%\caption{Monthly average $\text{PM}_{10}$ concentrations maps (upper plots) and relative width of the posterior interquartile range (lower plots).}
%\label{fig:mappeMensili_media}
%\end{figure}

\subsection{Model applications}

This section shows two potential applications of our model estimates for the assessment of air quality in Italy:  exceedance probability maps and population exposure to $\text{PM}_{10}$.

\paragraph{Exceedence}

To assess the risk of a pollutant, contaminated sites can be classified in terms of probabilities of exceeding (POE) a certain limit value \citep{Denby_et_al_2011}. For example, \citet{Yang_et_al_2016} show maps of probabilities of $\text{PM}_{2.5}$ concentrations exceeding 25 $\muup \text{g}/\text{m}^3$ for the Shandong Province (China). Similarly, in \cite{BLANGIARDO_2013} and \cite{Blangiardo_and_Cameletti_2015} the map of the posterior probability of exceeding the $\text{PM}_{10}$ threshold of 50 $\muup \text{g}/\text{m}^3$ is computed on a daily basis for Piemonte region (Italy).

POE maps represent a valid tool for those involved in managing the impacts of atmospheric pollution. The probability of exceeding a critical level in an area can be relevant both to increase public awareness in relation to air pollution, and to develop or improve mitigation actions on a local scale.

Based on the simulation results discussed in Section \ref{Sec:SpatialPrediction}, we calculated, for each cell of the reference grid, the probabilities of exceeding the daily limit value of 50 $\muup \text{g}/\text{m}^3$ for $\text{PM}_{10}$. Specifically, the exceedance probability of each cell was calculated as the number of exceedances divided by the total number of simulations (1000).

\begin{figure}%
    \centering
    \subfloat[January 26th]{\includegraphics[width=0.45\textwidth]{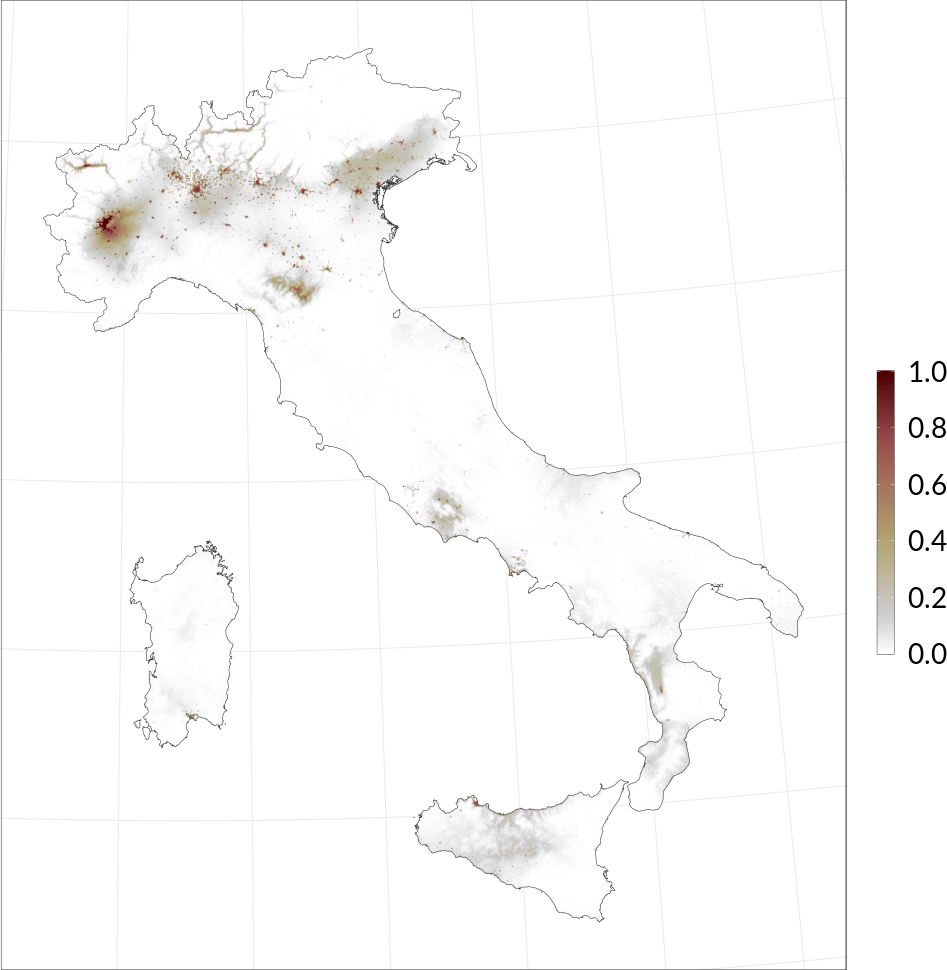}}%
    \qquad
    \subfloat[July 21st]{\includegraphics[width=0.45\textwidth]{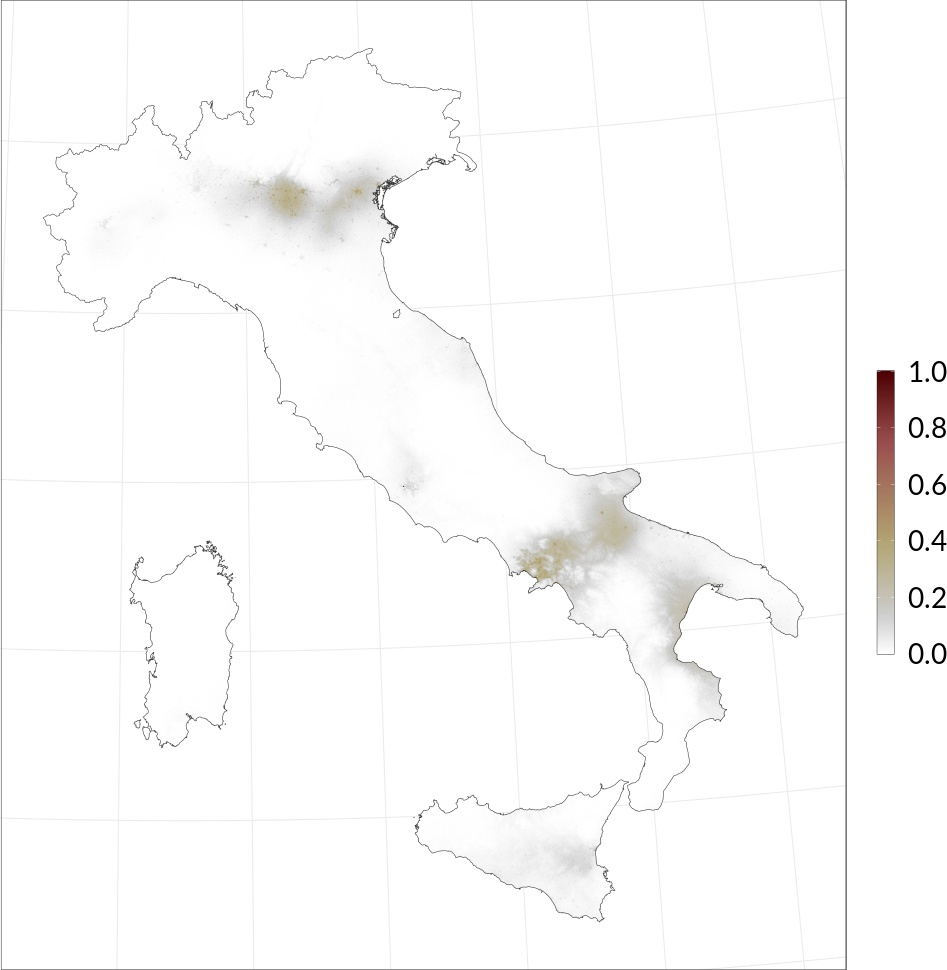}}%
    \caption{$\text{PM}_{10}$ exceedance probabilities (probabilities of $\text{PM}_{10}$ concentrations exceeding the threshold of 50 $\muup \text{g}/\text{m}^3$) for January 26th and July 21st, 2015.}%
    \label{fig:exceedence}%
\end{figure}

The final maps are shown in Figure \ref{fig:exceedence}. For the selected winter day (January 26th), the Po Valley exhibits several areas with high probabilities of exceedence, whose spatial distribution (around the large urban agglomerations) resembles, not surprisingly, the spatial pattern of the high pollutant concentrations seen in Figure \ref{fig:mappeGiornaliere_media} a). Conversely, the POE map for July 21st is characterized by low probability values (below 0.4), in accordance with the fact that $\text{PM}_{10}$ is not a critical pollutant in summer.

%\begin{figure} 
%\includegraphics[width=\textwidth]{images/probabilitaSuperamento.png}
%\caption{$\text{PM}_{10}$ exceedance probabilities (probabilities of $\text{PM}_{10}$ concentrations exceeding  $\muup \text{g}/\text{m}^3$.}
%\label{fig:exceedence}
%\end{figure}

\paragraph{Population exposure to $\text{PM}_{10}$}

The goal of many air pollution epidemiology studies is to estimate the effect of air pollution on health \citep{Sheppard_et_al_2005}. In this sense, comparing a limit value with the modeled concentrations is not sufficient for public health purposes, as it does no make any assumption about human exposition (the event of contact with a pollutant over a certain period of time) to air pollution \citep{Zou_et_al_a_2009}. 

Here, we combine the population density data and the model output concentrations to estimate the population exposure to $\text{PM}_{10}$ pollution in Italy at the municipality level. 
 
For the targeted municipality \textit{m}, the population-weighted $\text{PM}_{10}$ concentration level $e^m$ is given by:
 
 \begin{equation}\label{eq:exposure}
 e^m= \frac{ \sum_{i\in I_m} p_{i}\ c_{i}}{\sum_{i\in I_m} p_{i} } 
\end{equation}
where $I_m$ is the set of  of grid cells within  the administrative unit \textit{m}; $p_{i}$ and $c_{i}$ denote the population density  and $\text{PM}_{10}$ concentration level in the $i^{th}$ grid cell of \textit{m}, respectively.
 
\begin{figure}%
    \centering
    \subfloat[Annual mean concentrations]{\includegraphics[width=0.45\textwidth]{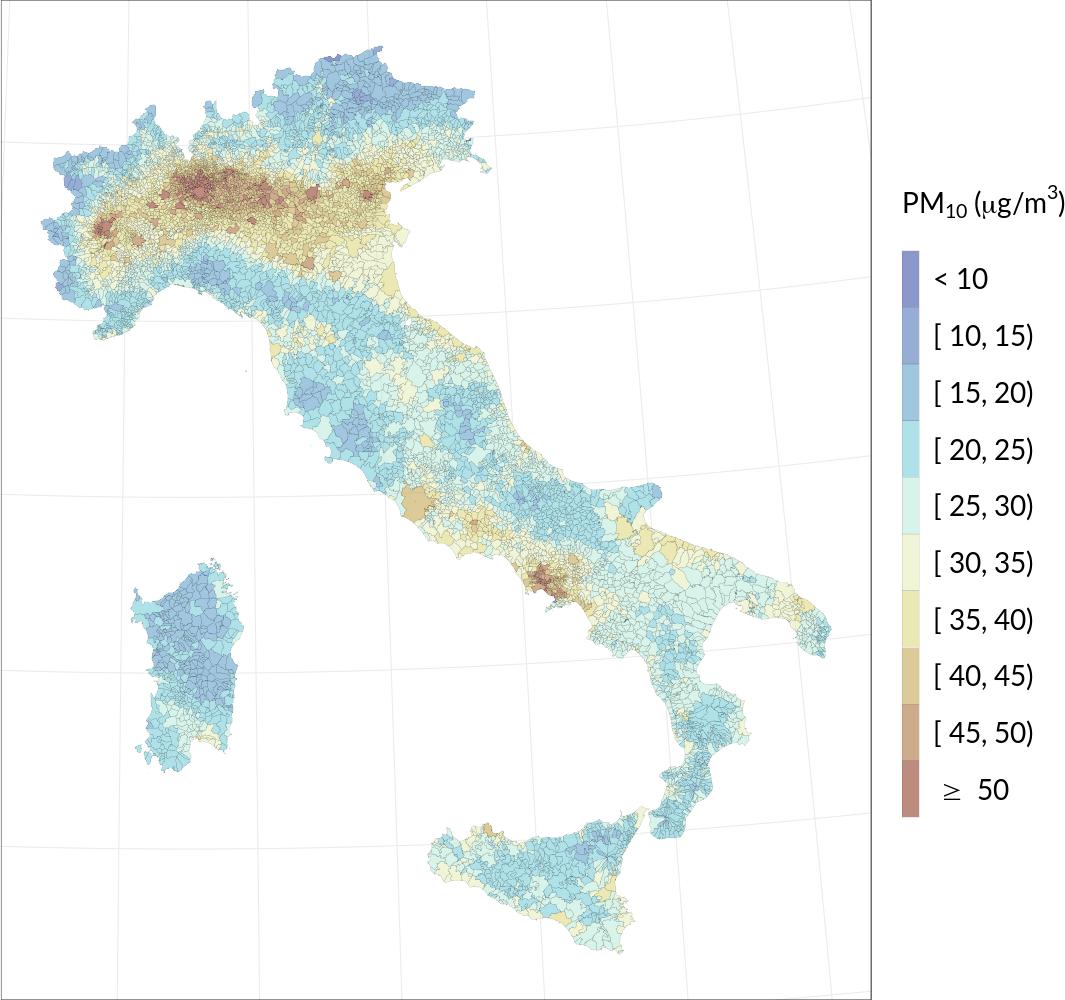}}%
    \qquad
    \subfloat[Annual 90.4 percentile concentrations]{\includegraphics[width=0.45\textwidth]{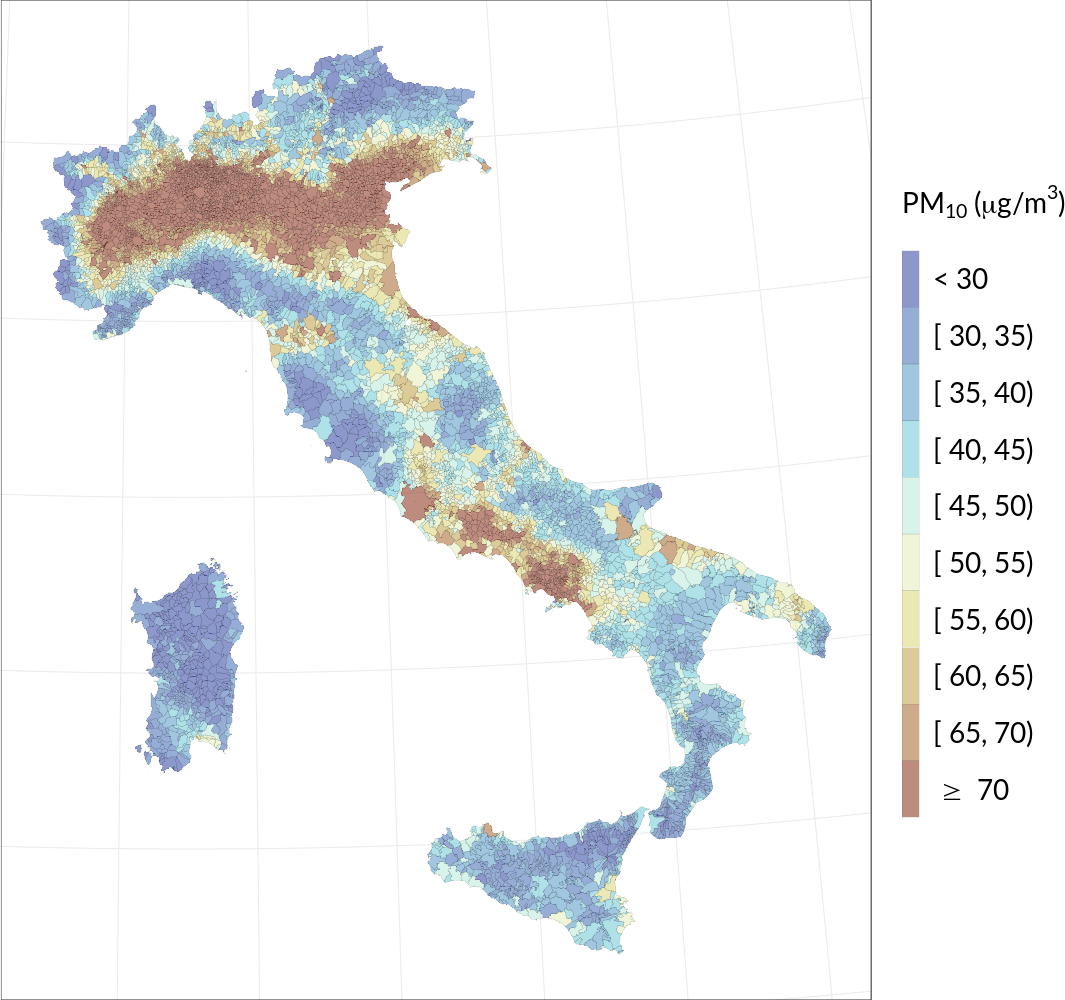}}%
    \qquad
    \subfloat[Annual 99.2 percentile concentrations]{\includegraphics[width=0.45\textwidth]{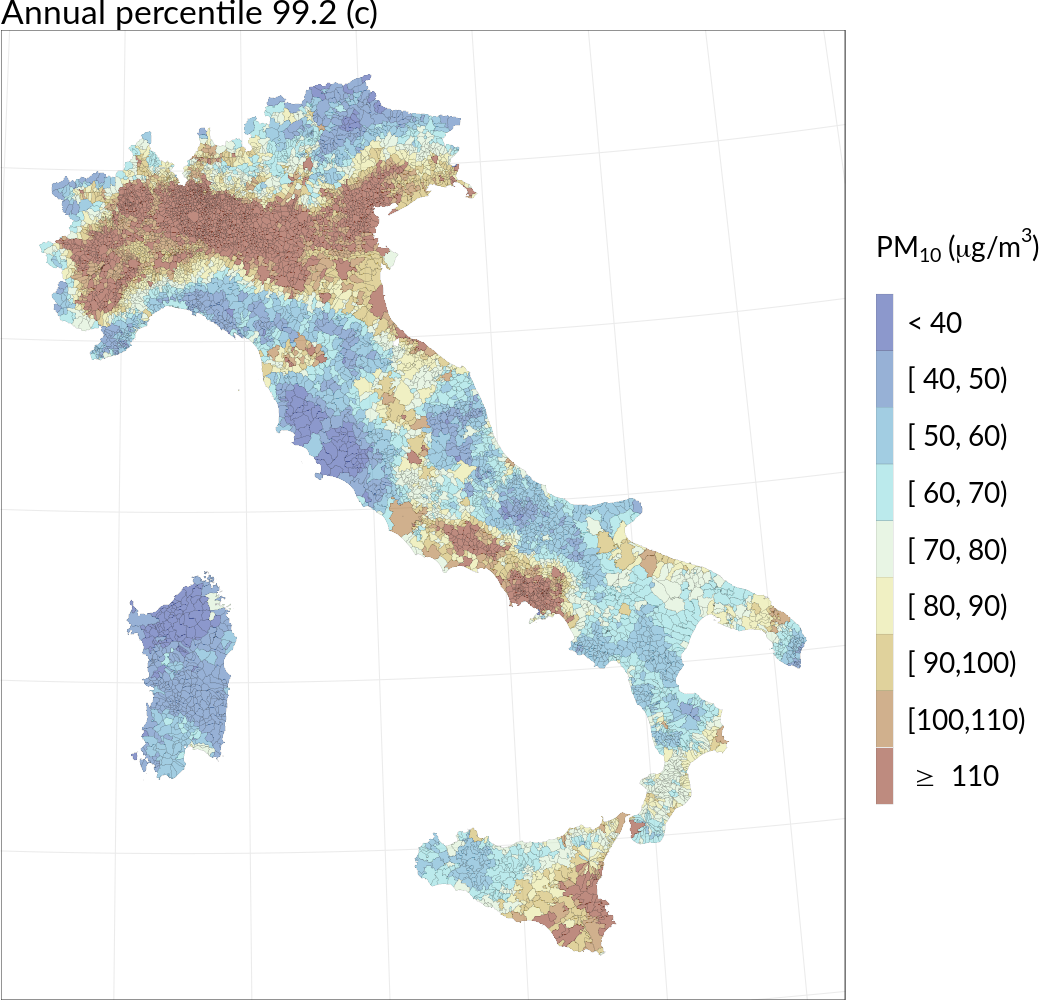}}

    \caption{Population exposure to $\text{PM}_{10}$ concentrations. EU air quality limit value for $\text{PM}_{10}$ daily concentrations is exceeded when 90.4 percentile is higher than 50 $\muup \text{g}/\text{m}^3$, while the more severe WHO air quality guideline is exceeded when 99.2 percentile is higher than 50 $\muup \text{g}/\text{m}^3$. The EU $\text{PM}_{10}$ annual average limit value is 40 $\muup \text{g}/\text{m}^3$, while the WHO air quality guideline is 20 $\muup \text{g}/\text{m}^3$.}%
    \label{fig:mappeEsposizione}%
\end{figure}

 For the considered case study, the $\text{PM}_{10}$ concentration levels $c_{i}$ are (a) the $\text{PM}_{10}$ annual mean concentrations, (b) the annual 90.4 percentile and (c) the annual 99.2 percentile, calculated  using the 365 daily interpolated surfaces discussed in Section \ref{Sec:SpatialPrediction}. For the population density data, we used the  national grid (1km $\times$ 1km) of the population density for 2011 of the Italian National Institute of Statistics (ISTAT, \url{https://www.istat.it/it/archivio/155162}). The final maps are displayed in Figure \ref{fig:mappeEsposizione}.
 
 The maps highlight the particular vulnerability to exposure to particulate pollution of the Po Basin, as well as the existence of other areas (the Sacco Valley and the Terni Basin in Central Italy, the agglomeration of Naples and Caserta in the south) where people are exposed to average levels above the WHO guidelines (20 $\muup \text{g}/\text{m}^3$ for the annual average) and the annual limit value settled by the European legislation (40 $\muup \text{g}/\text{m}^3$). The percentile maps (Figure \ref{fig:mappeEsposizione} b and c) indicate respectively the areas where the EU air quality limit value for $\text{PM}_{10}$ daily concentrations is exceeded (i.e., areas where the 90.4  percentile is higher than 50 $\muup \text{g}/\text{m}^3$), and the areas where the  more severe WHO air quality guideline for short-term exposure (24-hours) is exceeded  (99.2 annual percentile higher than 50 $\muup \text{g}/\text{m}^3$). The widespread exceedances of the air quality guidelines over the Italian territory arise the need to adopt more stringent policies to further reduce the anthropogenic emissions of PM and those of their precursors.
 
% We conclude this section observing that this kind of maps are relevant for epidemiological investigations on the effect of PM exposure on mortality/morbidity. In fact, considering that health outcomes and air pollution data are spatially misaligned \citep{CAMELETTI_2019}, it is necessary to obtain exposure estimates at the area level. 
 
%grafico mappe esposzione
%\begin{figure}
%\includegraphics[width=\textwidth]{images/esposizioneComuni_roma.png}
%\caption{Population exposure to $\text{PM}_{10}$ concentrations.}
%\label{fig:mappeEsposizione}
%\end{figure}

\section{Conclusions}\label{Sec:conclusion}

In this paper we proposed a Bayesian hierarchical spatio-temporal model for $\text{PM}_{10}$ daily concentrations. The model was applied, separately for each month, to the $\text{PM}_{10}$ concentrations measured during 2015 by the Italian monitoring network. This month-by-month approach represents an effective modeling solution for taking into account the seasonal variability of the phenomenon avoiding the use of a more complex year-based model which would require extremely higher computational costs. Moreover, with our modeling strategy it is possible to evaluate how the relationship between the considered predictors and $\text{PM}_{10}$ concentrations change across months. To the best of our knowledge no studies have assessed the predictors effect on the monthly timescale. From our results, we obtained that the covariates with the most pronounced seasonal effect are temperature, Planet Boundary Layer at 00:00, altitude and impervious surface. A clear but less marked impact of AOD on the $\text{PM}_{10}$ was also found.

The main outcome of our model is the continuous (1km $\times$ 1km) $\text{PM}_{10}$ map that we can estimate on a daily basis and is equipped with an uncertainty measure like the relative width of the posterior interquartile range. These high-resolution maps represent a fundamental tool for air quality management (at the national, regional and local level) with the aim of developing and monitoring programs and actions taken to improve air quality. As far as we know, there are very few other proposals in the statistical literature for this problem of mapping $\text{PM}_{10}$ concentrations on a large domain like Italy with a fine grid. In this regard, it is worth mentioning  \citet{Stafoggia_et_al_2017} and \citet{STAFOGGIA_2019}, which adopted LMM and a land-use random-forest model, respectively. Our opinion is that both the approaches are methodologically sound but are implemented by adopting a very complex modeling pipeline starting from missing data imputation and ending with predictions improvement by using small-scale predictors connected with very local sources. This gives rise to a computationally expensive modeling solution and to a difficulty in quantifying properly the uncertainty of the final predictions by taking into account all the variability sources. We believe that our modeling strategy, which is simple in its formulation and implementation, could represent a valid solution for this challenging problem which has an important connection with environment and human health protection. We would like to point out that, starting from the daily $\text{PM}_{10}$ maps, our modeling approach is also able to produce probability of exceedance and population-weighted exposure maps, that can be defined both at the grid or area level. While the former can be used to assess the compliance with air quality guidelines set for human health protection, the latter are necessary to link exposure to the health outcomes in epidemiological studies that investigate the long-term effect of air pollution exposure. 

The computational complexity of our analysis, given by the fact that we work with a large dataset (ca. 400 monitoring stations) and a fine spatio-temporal grid of about 11 millions cells, is managed by using the INLA-SPDE approach for model estimation and prediction. The cross-validation results suggest a good predictive performance of the model at almost all concentration levels, with the correlation between observed and predicted values ranging from 0.71 (in July) and 0.91 (in February), and the bias in the range 0.22 (August) - 1.07 $\muup \text{g}/\text{m}^3$ (January). Despite these encouraging results, large deviations between modeled and high extreme $\text{PM}_{10}$ observations remain an issue. This could be partly addressed in future work, for example, by improving the spatial resolution of the predictors (AOD and meteorological variables), including a quantitative description of the Saharan dust, or considering further sources of air pollution (fires, proximity to power plants, industrial facilities and so on). In this respect, we point out that originally our analysis considered a larger set of potential predictors, including those commonly used in PM modeling, such as the ``weekend effect'' or the Corine Land Cover land-use classification. However, most of them did not enter the final model because not statistically significant. Our final selection of predictors, including 11 variables, is supported by the analysis of the residuals of the models which appear to be uncorrelated both in space and time.

\section{Acknowledgments}

This research was partially funded by the Project “Piattaforma Tematica del Sentinel Collaborative GS per la
Qualità dell’Aria”. The agreement was signed between ASI (Agenzia Spaziale Italiana) and ISPRA (Istituto
Superiore per la Protezione e Ricerca Ambientale). CUP: F82F17000000005.

\bibliography{mybibfile}

\end{document}